\documentclass[prb,preprint,superscriptaddress,titlepage,showpacs,preprintnumbers,amsmath,amssymb,floats]{revtex4-2}

\usepackage{amsmath,mathtools}
\usepackage{natbib}
\usepackage{caption}
\usepackage{subcaption}
\usepackage{float}
\usepackage[pdfencoding=auto,colorlinks=true,linkcolor=blue]{hyperref}
\usepackage{amssymb}
\usepackage{gensymb}
\usepackage{amsthm}
\usepackage{graphicx}
\usepackage{physics}
\usepackage{ragged2e}
\usepackage {xcolor}

\newcommand{\bea}{\begin{eqnarray}}
\newcommand{\eea}{\end{eqnarray}}
\newcommand{\beq}{\begin{equation}}
\newcommand{\eeq}{\end{equation}}

\begin{document}  
\title{Magnetic field induced transition from nodeless to nodal superconductivity in $\beta$-PdBi$_{2}$}
\author{Emmanouil K. Kokkinis}
\email[]{kokki021@umn.edu}
\affiliation{School of Physics and Astronomy and William I. Fine Theoretical Physics Institute, University of Minnesota, Minneapolis, MN 55455, USA}
\author{Joseph J. Betouras}
\email[]{J.Betouras@lboro.ac.uk}
\affiliation{Department of Physics, Loughborough University, Loughborough LE11 3TU, England, United Kingdom}
\author{Andrey V. Chubukov}
\email[]{achubuko@umn.edu}
\affiliation{School of Physics and Astronomy and William I. Fine Theoretical Physics Institute, University of Minnesota, Minneapolis, MN 55455, USA}

\begin{abstract}
Recent tunneling measurements on $\beta$-PdBi$_2$ reported a magnetic field induced phase transition from a fully gapped to a nodal superconducting state. We develop a microscopic theory of superconductivity in this multi-band material, taking into account spin-orbit coupling and Zeeman splitting. We show that there are two attractive pairing channels in this system: an s-wave and a p-wave. At zero magnetic field, s-wave superconductivity wins. At a finite field, this channel becomes less favorable because Fermi surfaces split, and the gap symmetry changes to p-wave. We show that at a higher field, the excitation spectrum of a p-wave multi-band superconductor becomes gapless, with nodal points located in between the split Fermi surfaces. We argue that this behavior accounts for the experimentally observed field-induced nodal superconductivity in $\beta$-PdBi$_2$.
\end{abstract}
\maketitle
\section{Introduction}

Multiband superconductors with strong spin-orbit coupling (SOC) provide a fertile setting for unconventional pairing phenomena and magnetic field driven phase transitions \cite{Kuang_2021,Nadeem_2023,delaBarrera_2018,Lu_2015,Xi_2016,Saito_2016,Kuzmanovic_2022}. In such systems, spin-momentum locking and multiband electronic structure can modify the response of a superconducting state to a Zeeman field. While in a conventional superconductor a magnetic field reduces $T_c$ due to Zeeman splitting of Fermi surfaces and potentially introduces a finite-q superconducting order \cite{Fulde_1964,Larkin_1965}, more complex behavior can emerge when there are multiple pairing channels \cite{Fischer_2023,Yip_2014,Gorkov_2001}.

The layered compound $\beta$-PdBi$_2$ has recently attracted considerable attention in this context. This material has a layered crystal structure composed of covalently bonded trilayers, each consisting of a tetragonal Palladium (Pd) layer sandwiched between two AA-stacked tetragonal Bismuth (Bi) layers. Neighboring trilayers are held together by Van der Waals forces and are shifted from one another so that the Pd atoms in one trilayer are aligned with the Bi atoms in the other. Early studies established that $\beta$-PdBi$_2$ is a centrosymmetric superconductor with largely conventional thermodynamic properties \cite{Zhuravlev_1957,Imai_2012}, whereas later work revealed pronounced spin textures in its bulk and surface bands arising from strong SOC \cite{Shein_2013,Sakano_2015,Iwaya_2017,Xu_2019}. Most recently, tunneling spectroscopy measurements reported a field-induced evolution of the superconducting state: at low in-plane magnetic field  the data show a fully gapped STM spectrum \cite{Kacmarcik_2016,Biswas_2016,Herrera_2015,Powell_2025}, consistent with s-wave pairing, whereas above a critical field the spectrum develops a V-shape form and zero-bias conductance,  indicative of nodal quasiparticles \cite{Powell_2025}. Such behavior points to a change of pairing symmetry within the superconducting phase. This phenomenon is not expected within a single-channel BCS framework. 

In this work, we consider an effective low-energy two-dimensional model for a trilayer. This model describes itinerant electrons, originating from Bi p-orbitals and residing on the two Bi sublayers of each trilayer. These two sublayers define the layer (orbital) degree of freedom in our model.  In combination with a spin, this gives rise to  four electronic bands. We show that the observed behavior follows naturally from the competition between two pairing channels: s-wave and p-wave. In the  p-wave channel, the two fermions in a pair come from the same band. The p-wave momentum dependence of the interaction comes from the form factors from the diagonalization of the quadratic Hamiltonian in moving from orbital to band basis. The attraction in this channel is not present at the bare level, as attractive and repulsive pieces of the interaction exactly compensate each other, but emerges when we include the  renormalization of both pairing components from the particle-hole channel (the Kohn-Luttinger mechanism \cite{Kohn_1965}). We show that the dressed attractive interaction component becomes larger than the repulsive one. This holds at both zero and finite magnetic field. 
  
The pairing in the s-wave channel involves fermions from different bands.  At zero magnetic field, these bands are degenerate, so the Cooper logarithm is the same as if the two  fermions in a pair were from the same band. Here again the attractive and repulsive parts of the interaction exactly compensate each other at the bare level. There are two effects that lift this exact compensation. One is Kohn-Luttinger renormalization. We find that it acts against superconductivity as the dressed repulsive  s-wave  interaction component becomes larger than the attractive one.  This is consistent with the Kohn-Luttinger analysis for the Hubbard model ~\cite{Fay1968,Baranov1992}--an s-wave repulsion gets stronger after renormalization. The second effect is the electron-phonon interaction. It  favors s-wave pairing by either flipping the sign of both parts of the interaction or at least reducing the strength   of the repulsive pairing component.  We assume that the effect of electron-phonon interaction is stronger than Kohn-Luttinger-type renormalization and at zero magnetic field the attraction in the s-wave channel is stronger than in the p-wave channel. 
     
At a finite in-plane magnetic field, the bands from which fermions form an s-wave pair  are no longer degenerate.  This strongly suppresses s-wave pairing by the same reason as in a conventional BCS superconductor. The pairing in the p-wave channel, on the other hand,  remains robust.  As a result, increasing field induces a transition within a superconducting state from s-wave to p-wave gap symmetry.  We further demonstrate that  once the field exceeds a critical value,  the  p-wave gap function becomes nodal.   At a critical field, the minimum of the excitation energy reaches zero at four points, each located at momenta in between the original Fermi surfaces. Upon further increase of the field, each nodal point splits into two,  doubling the number of nodal points in the Bogoliubov quasiparticle spectrum.

This paper is organized as follows. In Sec. \ref{secII} we introduce an effective low-energy continuum model for $\beta$-PdBi$_2$. In Sec. \ref{secIIIA} we diagonalize the Hamiltonian at zero magnetic field. In Sec. \ref{secIIIB} we introduce a repulsive interaction and analyze superconductivity at zero field. In Sec. \ref{secIV} we derive the basis that diagonalizes the Hamiltonian at a finite magnetic field and re-analyze zero field superconductivity in this new basis. In Sec. \ref{secIVC} we study superconductivity at a finite field, analyze the Bogoliubov spectrum in the p-wave phase and establish the condition for the emergence of the nodal points. In Sec. \ref{secV} we compute the differential tunneling conductance and compare our results with experiment. Our conclusions are presented in Sec. \ref{secVI}.

\section{Model}
\label{secII}
The low-energy band structure of $\beta$-PdBi$_{2}$ around the $\Gamma$ point, where Fermi surfaces are located, can be effectively captured by a continuum model derived from a microscopic tight-binding description, where the relevant degrees of freedom arise from Bi p-orbitals that reside in the two sublayers of each trilayer. We use the effective low-energy continuum model for the band structure of $\beta$-PdBi$_{2}$, introduced in \cite{Powell_2025}. The model Hamiltonian reads:
\begin{equation}
\mathcal{H}_{0}=\frac{k^{2}}{2m}-\mu-\varepsilon\sigma_{x}+\alpha(k_{x}s_{y}-k_{y}s_{x})\sigma_{z}-hs_{x}
\end{equation}

\begin{figure}[H]
\begin{center}
\includegraphics[width=0.85\textwidth]{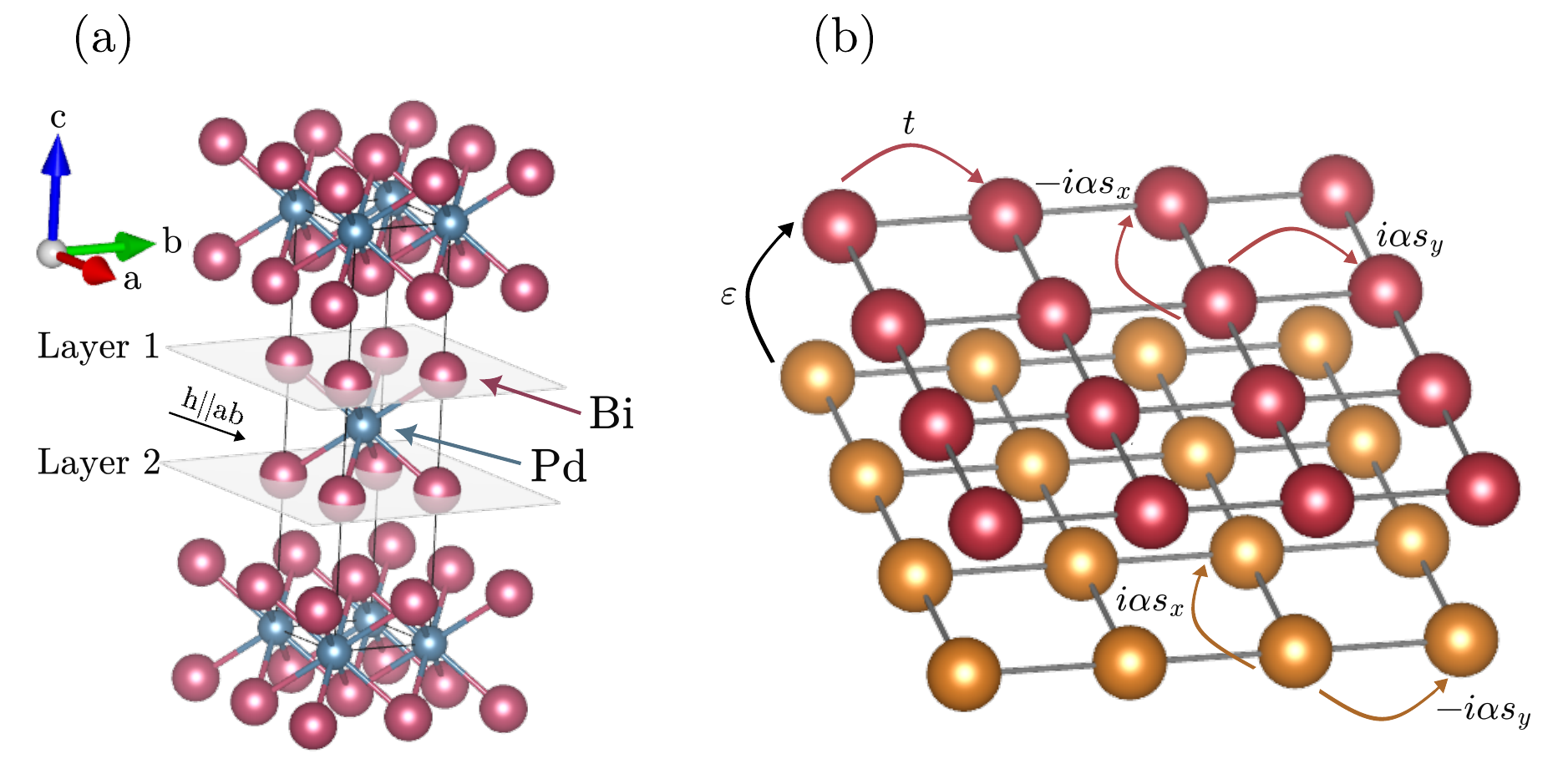}
\end{center}
\caption{(a) Crystal structure of $\beta$-PdBi$_2$. The two Bi layers within each trilayer (gray planes)  are related by inversion. The black arrow indicates the direction of the in-plane magnetic field. (b) Schematic representation of the terms in the Hamiltonian as hopping processes in real space. Red and orange atoms correspond to Bi atoms in layers 1 and 2, respectively. The Rashba spin-orbit coupling is represented as a direction-dependent spin-flip hopping in the two Bi layers.}
\label{fig:1}
\end{figure}
where $s_{i}$ and $\sigma_{i}$ are Pauli matrices corresponding
to spin and layer degrees of freedom respectively. Here, the term $\propto \varepsilon$ is the coupling between the two layers, the term $\propto \alpha$ is the Rashba spin-orbit coupling and $h$ is the transverse magnetic
field, which we direct along $x$. The model is depicted in Fig. \ref{fig:1}. For numerical calculations, we will use the parameters $\alpha=0.81 \;\text{eV}$, $m\equiv\frac{1}{2t}=-0.43\;\text{eV}^{-1}$ and $\varepsilon=0.63\;\text{eV}$. Those parameters are taken from \cite{Powell_2025}, where the authors determined them by comparing the bands obtained from this Hamiltonian with ARPES and DFT results \cite{Shein_2013,Sakano_2015,Wang_2017}, in order to reproduce the band structure around the $\Gamma$ point. A typical value for the chemical potential is $\mu=-2.22\;\text{eV}$. The exact value of $\mu$ is not qualitatively important for our analysis, the only requirement is that all four bands must cross the Fermi level.

\vspace{-1mm}
\section{Zero magnetic field}
\label{secIII}
\vspace{-1mm}
\subsection{Diagonalization of the non-interacting Hamiltonian}
\label{secIIIA}
\vspace{-1mm}
Let us first consider the case with $h=0$. The Hamiltonian in second-quantized notation is:
\begin{align}
\mathcal{H}_{0}= & \sum_{\textbf{k},s}\left(tk^{2}-\mu\right)a_{\textbf{k},s}^{\dagger}a_{\textbf{k},s}+\sum_{\textbf{k},s}\left(tk^{2}-\mu\right)b_{\textbf{k},s}^{\dagger}b_{\textbf{k},s}-\varepsilon\sum_{\textbf{k},s}\left(a_{\textbf{k},s}^{\dagger}b_{\textbf{k},s}+b_{\textbf{k},s}^{\dagger}a_{\textbf{k},s}\right)\nonumber \\
 & +i\alpha\sum_{\textbf{k},s}\left(ke^{i\varphi_{k}}a_{\textbf{k},\downarrow}^{\dagger}a_{\textbf{k},\uparrow}-ke^{-i\varphi_{k}}a_{\textbf{k},\uparrow}^{\dagger}a_{\textbf{k},\downarrow}\right)-i\alpha\sum_{\textbf{k},s}\left(ke^{i\varphi_{k}}b_{\textbf{k},\downarrow}^{\dagger}b_{\textbf{k},\uparrow}-ke^{-i\varphi_{k}}b_{\textbf{k},\uparrow}^{\dagger}b_{\textbf{k},\downarrow}\right)
\end{align}
where $\varphi_{k}=\text{Arg}\left[k_{x}+ik_{y}\right]$ and $a_{\textbf{k},s}$ and $b_{\textbf{k},s}$ are the fermionic operators for the
electrons in layer 1 and layer 2, respectively. It is convenient to diagonalize this Hamiltonian in two steps. First, with the transformation 
\begin{equation}
c_{\textbf{k},s}=\frac{a_{\textbf{k},s}+b_{\textbf{k},s}}{\sqrt{2}}
\end{equation}
\begin{equation}
d_{\textbf{k},s}=\frac{a_{\textbf{k},s}-b_{\textbf{k},s}}{\sqrt{2}}
\end{equation}
the Hamiltonian becomes Block diagonal:
\begin{equation}
\mathcal{H}_{0}=\sum_{\textbf{k}}\Psi_{\textbf{k}}^{\dagger}\begin{pmatrix}tk^{2}-\mu-\varepsilon & -i\alpha ke^{-i\varphi_{k}} & 0 & 0\\
i\alpha ke^{i\varphi_{k}} & tk^{2}-\mu+\varepsilon & 0 & 0\\
0 & 0 & tk^{2}-\mu-\varepsilon & i\alpha ke^{i\varphi_{k}}\\
0 & 0 & -i\alpha ke^{-i\varphi_{k}} & tk^{2}-\mu+\varepsilon
\end{pmatrix}\Psi_{\textbf{k}}
\end{equation}
where the column vector $\Psi_{\textbf{k}}=\begin{pmatrix}c_{\textbf{k},\uparrow} & d_{\textbf{k},\downarrow} & c_{\textbf{k},\downarrow} & d_{\textbf{k},\uparrow}\end{pmatrix}^{T}$. Subsequently, the transformation that fully diagonalizes $\mathcal{H}_{0}$ is:
\begin{align}
c_{\textbf{k},\uparrow}&=\gamma_{k}\cdot e_{\textbf{k}}+ie^{-i\varphi_{k}}z_{k}\cdot f_{\textbf{k}}\nonumber\\
d_{\textbf{k},\downarrow}&=ie^{i\varphi_{k}}z_{k}\cdot e_{\textbf{k}}+\gamma_{k}\cdot f_{\textbf{k}}\nonumber\\
c_{\textbf{k},\downarrow}&=\gamma_{k}\cdot\bar{e}_{\textbf{k}}-ie^{i\varphi_{k}}z_{k}\cdot\bar{f}_{\textbf{k}}\nonumber\\
d_{\textbf{k},\uparrow}&=-ie^{-i\varphi_{k}}z_{k}\cdot\bar{e}_{\textbf{k}}+\gamma_{k}\cdot\bar{f}_{\textbf{k}}\nonumber\\
\begin{Bmatrix}\gamma_k\\
z_k
\end{Bmatrix}&=\frac{1}{\sqrt{2}}\left[1\mp\frac{\varepsilon}{\sqrt{\varepsilon^{2}+\alpha^{2}k^{2}}}\right]^{1/2}
\label{eq:6}
\end{align}
The new fermionic operators $e_{\textbf{k}}$, $\bar{e}_{\textbf{k}}$ and $f_{\textbf{k}}$, $\bar{f}_{\textbf{k}}$ are two pairs of degenerate doublets. The operators in each pair can be viewed as components of a pseudospin, due to the spin-momentum locking arising from the SOC. The diagonalized Hamiltonian is expressed as 
\begin{equation}
\mathcal{H}_{0}=\sum_{\textbf{k}}\varepsilon_{+}\left(k\right)\left[e_{\textbf{k}}^{\dagger}
e_{\textbf{k}}+\bar{e}_{\textbf{k}}^{\dagger}\bar{e}_{\textbf{k}}\right]+\sum_{\textbf{k}}
\varepsilon_{-}\left(k\right)\left[f_{\textbf{k}}^{\dagger}f_{\textbf{k}}+\bar{f}_{\textbf{k}}^{\dagger}\bar{f}_{\textbf{k}}\right]
\label{aa:1}
\end{equation}
with the two dispersions given by
\begin{equation}
\varepsilon_{\pm}\left(k\right)=tk^{2}-\mu\pm\sqrt{\varepsilon^{2}+\alpha^{2}k^{2}}
\end{equation} 
The corresponding Fermi momenta $k_{F \pm}$ are 
\begin{equation}
k_{F\pm}=\left[\frac{\alpha^2+2\mu t\pm\sqrt{(\alpha^2+2\mu t)^2-4t^2(\mu^2-\varepsilon^2)}}{2t^2}\right]^{1/2}
\end{equation}

\begin{figure}[H]
\begin{minipage}{.5\textwidth}
    \centering
    \includegraphics[scale=0.50]{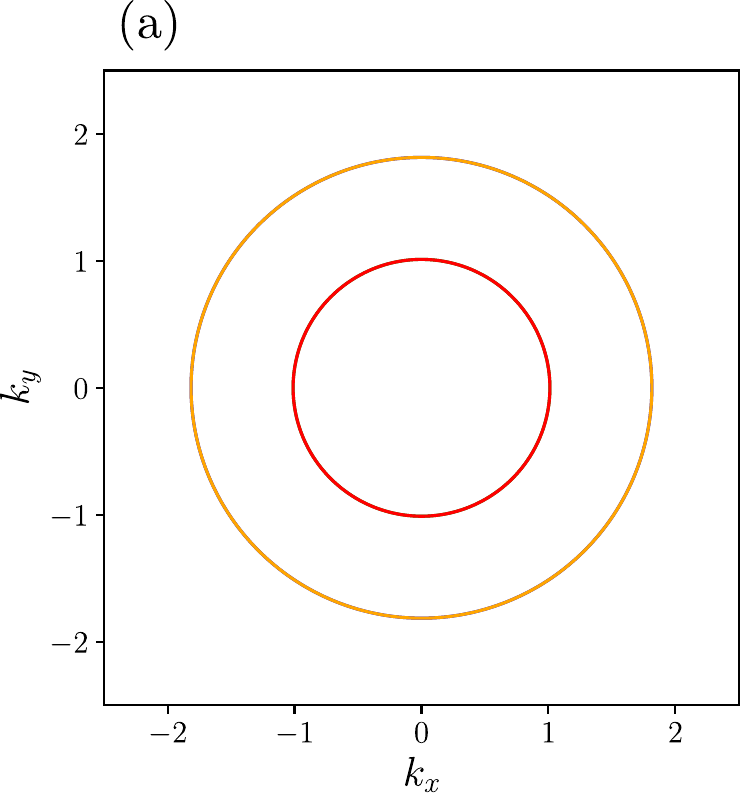}
    \phantomsubcaption\label{fig:2a}
\end{minipage}
\begin{minipage}{.5\textwidth}
    \centering
    \includegraphics[scale=0.50]{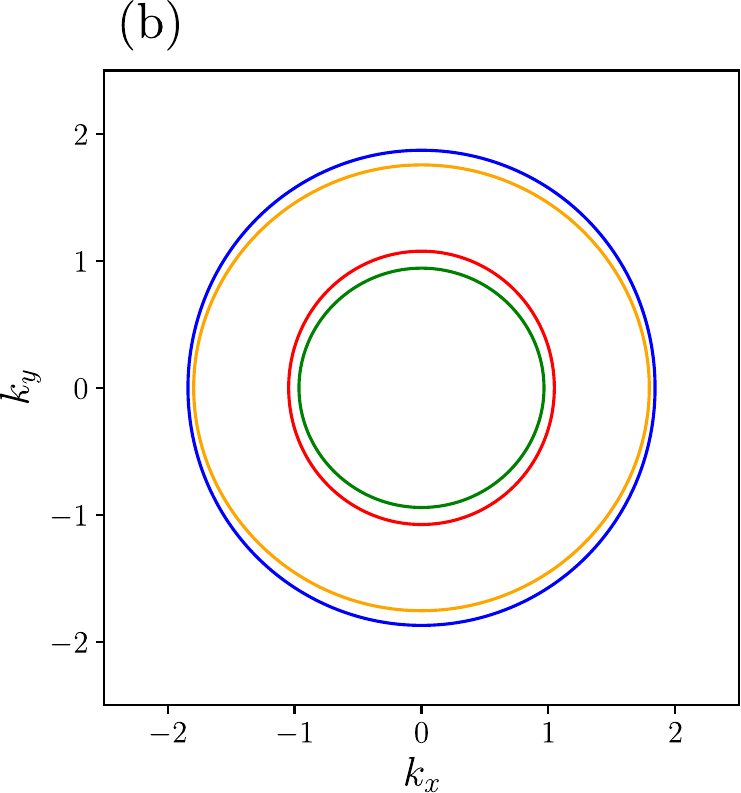}
    \phantomsubcaption\label{fig:2b}
\end{minipage}
\caption{(a) Fermi surfaces in the normal state. (a) $h=0$. Each of the two bands is doubly degenerate. (b)  $h\neq0$ (schematic).}
\label{fig:2}
\end{figure}
Eq. (\ref{aa:1}) describes two doubly degenerate hole bands with one smaller and one larger spherically symmetric Fermi surfaces, as illustrated in Fig. \ref{fig:2a}.  Each  dispersion mixes spin-up and spin-down states because spin is not a good quantum number in the presence of SOC. Because  each band is doubly degenerate,  the transformation  we have chosen  to diagonalize the quadratic form is not  unique: any unitary transformation that mixes the fermions $e_{\textbf{k}}$ with $\bar{e}_{\textbf{k}}$, and $f_{\textbf{k}}$ with $\bar{f}_{\textbf{k}}$ keeps $\mathcal{H}_{0}$ intact. The one that we have chosen is the most convenient  for the study of the two different superconducting  states at zero field.
\subsection{Superconductivity at h=0}
\label{secIIIB}
To study superconductivity, we introduce a repulsive four-fermion density-density interaction. In the original basis,  it is 
\begin{align}
\mathcal{H}_{int}&=U\sum_{\substack{\mathbf{k},\mathbf{p},\mathbf{q} \\ s_{1},s_{2}}}\left[a_{\textbf{k},s_{1}}^{\dagger}a_{\textbf{p},s_{1}}a_{-\textbf{k}+\textbf{q},s_{2}}^{\dagger}a_{-\textbf{p}+\textbf{q},s_{2}}+b_{\textbf{k},s_{1}}^{\dagger}b_{\textbf{p},s_{1}}b_{-\textbf{k}+\textbf{q},s_{2}}^{\dagger}b_{-\textbf{p}+\textbf{q},s_{2}}\right]\nonumber \\
&+2V\sum_{\substack{\mathbf{k},\mathbf{p},\mathbf{q} \\ s_{1},s_{2}}}a_{\textbf{k},s_{1}}^{\dagger}a_{\textbf{p},s_{1}}b_{-\textbf{k}+\textbf{q},s_{2}}^{\dagger}b_{-\textbf{p}+\textbf{q},s_{2}}
\end{align}
where $U$ is the intralayer and $V$ is the interlayer repulsion. In the $e$-$f$  basis, the interaction Hamiltonian projected into the BCS  channel conveniently decomposes into two terms (see Appendix \ref{AppendixA}), which describe two distinct superconducting orders. Namely, $\mathcal{H}_{int} = \mathcal{H}_{int}^{\text{intra}} + \mathcal{H}_{int}^{\text{inter}}$, where
\begin{align}
\mathcal{H}_{int}^{\text{intra}}&=\sum_{\textbf{k},\textbf{p}}\gamma_k\gamma_pz_kz_p\Bigl\{\left(U+V\right)\left[B_{\textbf{k}}^\dagger B_{\textbf{p}}+\bar{B}_{\textbf{k}}^\dagger \bar{B}_{\textbf{p}}\right]+\left(U-V\right)\left[B_{\textbf{k}}^\dagger \bar{B}_{\textbf{p}}+\bar{B}_{\textbf{k}}^\dagger B_{\textbf{p}}\right]\Bigr\}\nonumber\\
\mathcal{H}_{int}^{\text{inter}} & =\sum_{\textbf{k},\textbf{p}}\Bigl\{(U+V)S_{\textbf{kp}}+(U-V)\left[1-S_{\textbf{kp}}\right]\Bigr\}\left[e_{\textbf{k}}^{\dagger}\bar{e}_{-\textbf{k}}^{\dagger}\bar{e}_{-\textbf{p}}e_{\textbf{p}}+f_{\textbf{k}}^{\dagger}\bar{f}_{-\textbf{k}}^{\dagger}\bar{f}_{-\textbf{p}}f_{\textbf{p}}\right] \nonumber\\
 & -\sum_{\textbf{k},\textbf{p}}\Bigl\{(U-V)S_{\textbf{kp}}+(U+V)\left[1-S_{\textbf{kp}}\right]\Bigr\}
 \left[e_{\textbf{k}}^{\dagger}\bar{e}_{-\textbf{k}}^{\dagger}\bar{f}_{-\textbf{p}}f_{\textbf{p}}+
f^{\dagger}_{\textbf{k}}\bar{f}^{\dagger}_{-\textbf{k}}   \bar{e}_{-\textbf{p}}e_{\textbf{p}}\right]
\label{eq:10}
\end{align}
and we defined
\begin{align}
B_{\textbf{k}}^\dagger&=e^{-i\varphi_k}e_{\textbf{k}}^\dagger e_{-\textbf{k}}^\dagger-e^{i\varphi_k}f_{\textbf{k}}^\dagger f_{-\textbf{k}}^\dagger\nonumber \\
\bar{B}_{\textbf{k}}^\dagger&=e^{i\varphi_k}\bar{e}_{\textbf{k}}^\dagger \bar{e}_{-\textbf{k}}^\dagger-e^{-i\varphi_k}\bar{f}_{\textbf{k}}^\dagger \bar{f}_{-\textbf{k}}^\dagger\nonumber \\
S_{\textbf{kp}}&=(\gamma_k \gamma_p)^2+(z_k z_p)^2
\end{align}
There are six possible condensates with zero total momentum: four intraband $\Delta_{ee}(\textbf{k})\sim\left\langle e_{-\textbf{k}}e_\textbf{k}\right\rangle$, $\Delta_{\bar{e}\bar{e}}(\textbf{k})\sim\left\langle \bar{e}_{-\textbf{k}}\bar{e}_\textbf{k}\right\rangle$,  $\Delta_{ff}(\textbf{k})\sim\left\langle f_{-\textbf{k}}f_\textbf{k}\right\rangle$, $\Delta_{\bar{f}\bar{f}}(\textbf{k})\sim\left\langle \bar{f}_{-\textbf{k}}\bar{f}_\textbf{k}\right\rangle$, and two interband $\Delta_{e\bar{e}}(\textbf{k})\sim\left\langle \bar{e}_{-\textbf{k}}{e}_\textbf{k}\right\rangle$ and $\Delta_{f\bar{f}}(\textbf{k})\sim\left\langle \bar{f}_{-\textbf{k}}f_\textbf{k}\right\rangle$. The four intraband components $\Delta_{ee}(\textbf{k})$, $\Delta_{\bar{e}\bar{e}}(\textbf{k})$, $\Delta_{ff}(\textbf{k})$ and $\Delta_{\bar{f}\bar{f}}(\textbf{k})$ are coupled by $\mathcal{H}_{int}^{\text{intra}}$. The two interband components $\Delta_{e\bar{e}}\left(\textbf{k}\right)$ and $\Delta_{f\bar{f}}\left(\textbf{k}\right)$ are coupled by $\mathcal{H}_{int}^{\text{inter}}$. We study intraband and interband pairing separately.

\subsubsection{Intraband paring (p-wave channel)}

The set of coupled linearized equations for the superconducting gaps $\Delta_{\alpha\alpha}(\textbf{k})$ is given by
\begin{align}
\Delta_{\alpha\alpha}\left(\textbf{k}\right)=-\sum_{\beta} \intop\frac{d^{2}\textbf{p}}{\left(2\pi\right)^{2}}\frac{U^{\text{intra}}_{\alpha\beta}(\textbf{k},\textbf{p})\Delta_{\beta\beta}\left(\textbf{p}\right)}{\varepsilon_{\beta}(\textbf{p})}\tanh\left(\frac{\varepsilon_{\beta}(\textbf{p})}{2T_c}\right)
\label{eq:3.4}
\end{align}
where $\alpha$ and $\beta$ run over $e$, $\bar{e}$, $f$, $\bar{f}$ and the components of $U^{\text{intra}}_{\alpha\beta}(\textbf{k},\textbf{p})$ are the sixteen terms in $\mathcal{H}_{int}^{\text{intra}}$ in (\ref{eq:10}).  One can easily verify that the solution of (\ref{eq:3.4}) has the form 
\begin{align}
\Delta_{ee}(\textbf{k}) & =\Delta_{ee}\gamma_{k}z_{k}e^{-i\varphi_{k}} \nonumber \\
\Delta_{\bar{e}\bar{e}}(\textbf{k}) & =\Delta_{\bar{e}\bar{e}}\gamma_{k}z_{k}e^{i\varphi_{k}} \nonumber \\
\Delta_{ff}(\textbf{k}) & =\Delta_{ff}\gamma_{k}z_{k}e^{i\varphi_{k}} \nonumber \\
\Delta_{\bar{f}\bar{f}}(\textbf{k}) & =\Delta_{\bar{f}\bar{f}}\gamma_{k}z_{k}e^{-i\varphi_{k}} 
\label{qq_1}
\end{align}
where we recall that $\varphi_{k}=\text{Arg}\left[k_{x}+ik_{y}\right]$. We immediately see that the 
superconducting gaps have a structure of $p_x\pm ip_y$ and change sign under $\textbf{k}\rightarrow-\textbf{k}$. The p-wave structure of the gaps originates from the $e^{\pm i\varphi_k}$ form factors in $\mathcal{H}_{int}^{\text{intra}}$. However, time reversal has not been broken. This can be seen from the fact that under complex conjugation and $e_{\textbf{k}}\leftrightarrow\bar{e}_{\textbf{k}}$, $f_{\textbf{k}}\leftrightarrow\bar{f}_{\textbf{k}}$, the gap structure remains invariant. One can easily verify that the linearized gap equation can be re-expressed as a system of algebraic equations:
\begin{equation}
\begin{pmatrix}
\Delta_{{ee}}+\Delta_{\bar{e}\bar{e}} \\
\Delta_{ff}+\Delta_{\bar{f}\bar{f}} \\
\Delta_{{ee}}-\Delta_{\bar{e}\bar{e}} \\
\Delta_{ff}-\Delta_{\bar{f}\bar{f}}
\end{pmatrix}
=
\begin{pmatrix}
-UL_+ & UL_- & 0 & 0 \\
UL_+ & -UL_- & 0 & 0 \\
0 & 0 & -VL_+ & VL_- \\
0 & 0 & VL_+ & -VL_-
\end{pmatrix}
\begin{pmatrix}
\Delta_{{ee}}+\Delta_{\bar{e}\bar{e}} \\
\Delta_{ff}+\Delta_{\bar{f}\bar{f}} \\
\Delta_{{ee}}-\Delta_{\bar{e}\bar{e}} \\
\Delta_{ff}-\Delta_{\bar{f}\bar{f}}
\end{pmatrix}
\label{eq:14}
\end{equation}
where we have defined 
\begin{equation}
L_{\pm}=2\intop\frac{d^{2}\textbf{p}}{\left(2\pi\right)^{2}}\frac{\left(\gamma_{p}z_{p}\right)^{2}}{\varepsilon_{\pm}\left(p\right)}\tanh\left(\frac{\varepsilon_{\pm}\left(p\right)}{2T_{c}}\right)\approx N_{F\pm}\frac{\alpha^2k_{F\pm}^2}{\varepsilon^2+\alpha^2k_{F\pm}^2}\log\frac{1.13\omega_D}{T_c}
\label{eq:3.6}
\end{equation}

Here, $N_{F\pm}$ denotes the density of states at the Fermi level associated with the bands $\varepsilon_{\pm}(k)$. In evaluating the integral in Eq. (\ref{eq:3.6}), we restrict the energy integration to a narrow window $\left[-\omega_D,\omega_D\right]$ around the Fermi surface. Within this window, all non-singular quantities are approximated as constants and taken outside of the momentum integral in (\ref{eq:3.6}). Number-wise, the densities of states are $N_{F+}=0.083\;\text{eV}^{-1}$, $N_{F-}=0.054\;\text{eV}^{-1}$. We see from Eq. (\ref{eq:14})  that the linearized gap equation decomposes into two different blocks in which the gaps on the two degenerate bands develop either with the same or the opposite sign. On general grounds, we expect  intra-layer $U$ to be larger than inter-layer $V$. We consider this case here and for completeness discuss  the case $V > U$ in Appendix \ref{AppendixD}.  For $U >V$, one can explicitly verify that $\Delta_{\bar{e}\bar{e}}=\Delta_{{ee}}$ and $\Delta_{\bar{f}\bar{f}}=\Delta_{ff}$. The functions $\Delta_{{ee}}$  and $\Delta_{{ff}}$  satisfy the 2x2 set of equations 
\begin{equation}
\begin{pmatrix}
\Delta_{{ee}} \\
\Delta_{{ff}}
\end{pmatrix}
=
\begin{pmatrix}
-UL_+ & UL_- \\
UL_+ & -UL_-
\end{pmatrix}
\begin{pmatrix}
\Delta_{{ee}} \\
\Delta_{{ff}}
\end{pmatrix}
\label{eq:17}
\end{equation}
Note that the interlayer interaction $V$ drops out of the gap equation. Analyzing (\ref{eq:17}), we see that there are repulsive (intra-pocket) and attractive (inter-pocket) components of the interaction of equal magnitude $U$. Evaluating the determinant, we find that it vanishes, i.e., the attractive and repulsive parts of the interaction compensate each other. However, this degeneracy breaks down once we treat each interaction as irreducible in the pairing channel and dress it up by the renormalization from the particle-hole channel (they are often called Kohn-Luttinger renormalizations \cite{Kohn_1965}). The irreducible repulsive and attractive components are generally non-equal, and we label them as $U_1$ and $U_2$. The computation of $U_2$ and $U_1$ to second order in the interaction is straightforward but cumbersome because of coherence factors $\gamma_k$ and $z_k$. We discuss the details in Appendix \ref{AppendixB} and here cite the result: the dressed $U_2$ is larger then the dressed $U_1$. For our input parameters,
\begin{equation}
U_1 = U\left[1 + 1.985N_FU\right], ~~ U_2 = U\left[1 + 2.206N_FU\right]
\end{equation}
where we have defined $N_F=\frac{N_{F+}+N_{F-}}{2}$. Solving Eq. (\ref{eq:14}) with $U_2>U_1$, we find the equation on $T_c$ in the form
\begin{equation}
U_2^2-U_1^2=\frac{1+U_1(L_++L_-)}{L_+L_-}
\label{eq:Tc1}
\end{equation}  

The solution is a finite $T_c$. For $U_2 -U_1 \ll U_1$, as in our case, $\log{\omega_D/T_c}\propto 1/(U_2-U_1)=4.5/(U N_F)^2$. This is similar to the cases of cuprates and iron-based superconductors, where the Kohn-Luttinger mechanism also enhances inter-pocket/inter-patch interaction compared to intra-pocket/intra-patch one \cite{Kuroki_2009,Chubukov_2012,Maiti_2014}, but with one distinction: in our case the inter-pocket interaction is attractive, hence   the gap functions $\Delta_{ee}$ and $\Delta_{ff}$ have the same sign. The attraction comes from the coherence factors after the quadratic Hamiltonian is diagonalized. We assume and then verify by comparing with the experimental data that the relevant $U \sim 1-2\;\text{eV}$. For such $U$, $U N_{F \pm}$ are small, i.e., the pairing problem falls into weak coupling range. As in the case of all Kohn-Luttinger-type calculations at weak coupling, Eq. (\ref{eq:Tc1}) proves that $T_c$ is non-zero, but to get its actual value, one would need to compute renormalizations to order $U^4$ as these terms modify the expression for $T_c$ in (\ref{eq:Tc1}) to $\log{\omega_D/T_c}\propto 1/(U_2-U_1)  = (4.5/(UN_F)^2) (1 + a U N_F + b (U N_F)^2)$, and  the values of both  $a$ and $b$ (the prefactors for the $U^3$ and $U^4$ terms) are relevant for $T_c$ (see e.g., Ref \cite{Baranov1992}). For this reason, below we keep Eq. (\ref{eq:Tc1}) without higher-order corrections, but treat $U_1$ and $U_2 > U_1$ as phenomenological parameters, which we choose to fit the experimental data. Then solving the non-linear gap equation 
\begin{align}
\Delta_{\alpha\alpha}=-\sum_{\beta}U^{\text{intra}}_{\alpha\beta}\Delta_{\beta\beta} \intop\frac{d^{2}\textbf{p}}{\left(2\pi\right)^{2}}\frac{(\gamma_pz_p)^2}{E_{\beta}(\textbf{p})}\tanh\left(\frac{E_{\beta}(\textbf{p})}{2T}\right)
\end{align}
\begin{figure}[H]
\begin{center}
\includegraphics[width=.6\textwidth]{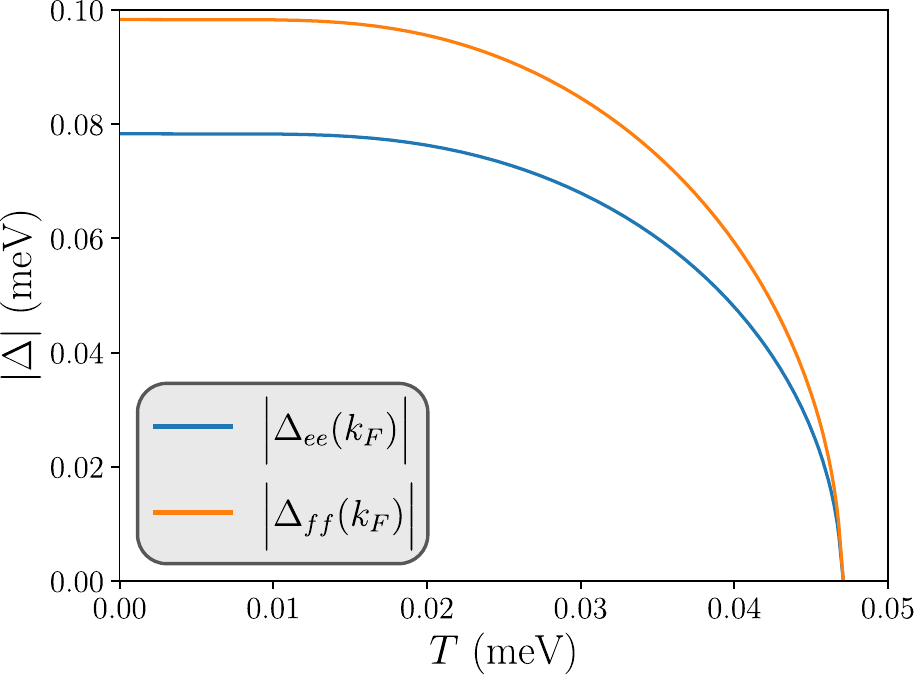}
\end{center}
\caption{Superconducting gaps $|\Delta_{ee}(k_F)|$ (blue line) and $|\Delta_{ff}(k_F)|$ (orange line) evaluated at the corresponding Fermi surface as functions of the temperature. Those are obtained from solving the non-linear gap equation. The modulus of each gap has no angular dependence, although the gap symmetry is p-wave. We used $U_1=0$, $U_2=2.6$ eV and $\omega_D=0.1$ eV.}
\label{fig:3}
\end{figure}
where $E_{\beta}(\textbf{k})=\sqrt{(\varepsilon_{\beta}\left(k\right))^2+\gamma_k^2z_k^2|\Delta_{\beta\beta}|^2}$, and $U^{\text{intra}}_{\alpha\beta}=U_1$ for fermions within the same pocket and $U^{\text{intra}}_{\alpha\beta}=-U_2$ for fermions from different pockets, we obtain variations of $\Delta_{ee}$ and $\Delta_{ff}$ with temperature. In Fig. \ref{fig:3} we show the solution for a representative  value of $U_{1,2}$, chosen to match the experimentally detected p-wave gap $|\Delta_p| \leq  0.1\; \text{meV}$, detected in STM at $B^{||}\gtrsim0.2$ T (Ref. \cite{Powell_2025}).

\subsubsection{Interband pairing (s-wave channel)}
The linearized gap equation in the interband channel couples   $\Delta_{e{\bar e}} ({\bf k}_F)$  and $\Delta_{f{\bar f}} ({\bf k}_F)$:
\begin{equation}
\Delta_{\alpha\bar{\alpha}}\left(\textbf{k}\right)=-\sum_{\beta} \intop\frac{d^{2}\textbf{p}}{\left(2\pi\right)^{2}}\frac{U^{\text{inter}}_{\alpha\bar{\alpha};\bar{\beta}\beta}(\textbf{k},\textbf{p})\Delta_{\beta\bar{\beta}}\left(\textbf{p}\right)}{2\varepsilon_{\beta}(\textbf{p})}\tanh\left(\frac{\varepsilon_{\beta}(\textbf{p})}{2T_c}\right)
\end{equation}
where $\alpha$ and $\beta$ are  $e$ or $f$ and $U^{\text{inter}}_{\alpha\bar{\alpha};\bar{\beta}\beta}(\textbf{k},\textbf{p})$ are the four terms in $\mathcal{H}_{int}^{\text{inter}}$. This $2 \times 2$ set  of integral equations  reduces to a set of four algebraic equations by using the ansatz $\Delta_{e\bar{e}}(\textbf{k})=\Delta_1 \gamma_k^2+\Delta_2z_k^2$ and $\Delta_{f\bar{f}}(\textbf{k})=\Delta_3 \gamma_k^2+\Delta_4z_k^2$.  There is no dependence on the polar angle $\varphi$, hence  the pairing symmetry is s-wave. The Fermi surfaces for $e_{\textbf{k}}$ and ${\bar e_{\textbf{k}}}$ fermions are degenerate and the same  holds for $f_{\textbf{k}}$ and ${\bar f_{\textbf{k}}}$ fermions. Hence, $e_\textbf{k}$ and ${\bar e}_{-\textbf{k}}$ fermions  and   $f_\textbf{k}$ and ${\bar f}_{-\textbf{k}}$  fermions can simultaneously be placed on the Fermi surface. As a consequence, the Cooper logarithm is present. The analysis of the conditions for the existence of $T_c$ is more involved here because both $U$ and $V$ terms  contribute to the pairing. We present the full treatment in Appendix \ref{AppendixC} and here consider the case $V=0$ for direct comparison with the p-wave pairing. The functions $\Delta_{e\bar{e}}$ and $\Delta_{f\bar{f}}$ satisfy the 2x2 set of equations
\begin{equation}
\begin{pmatrix}
\Delta_{{e\bar{e}}} \\
\Delta_{{f\bar{f}}}
\end{pmatrix}
=
\begin{pmatrix}
-U\tilde{L}_+ & U\tilde{L}_- \\
U\tilde{L}_+ & -U\tilde{L}_-
\end{pmatrix}
\begin{pmatrix}
\Delta_{{e\bar{e}}} \\
\Delta_{{f\bar{f}}}
\end{pmatrix}
\label{eq:22}
\end{equation}
where in this case
\begin{equation}
\tilde{L}_{\pm}=\frac{1}{2}\intop\frac{d^{2}\textbf{p}}{\left(2\pi\right)^{2}}\frac{1}{\varepsilon_{\pm}\left(p\right)}\tanh\left(\frac{\varepsilon_{\pm}\left(p\right)}{2T_{c}}\right)\approx N_{F\pm}\log\frac{1.13\omega_D}{T_c}
\end{equation}

\begin{figure}[H]
\begin{center}
\includegraphics[width=.6\textwidth]{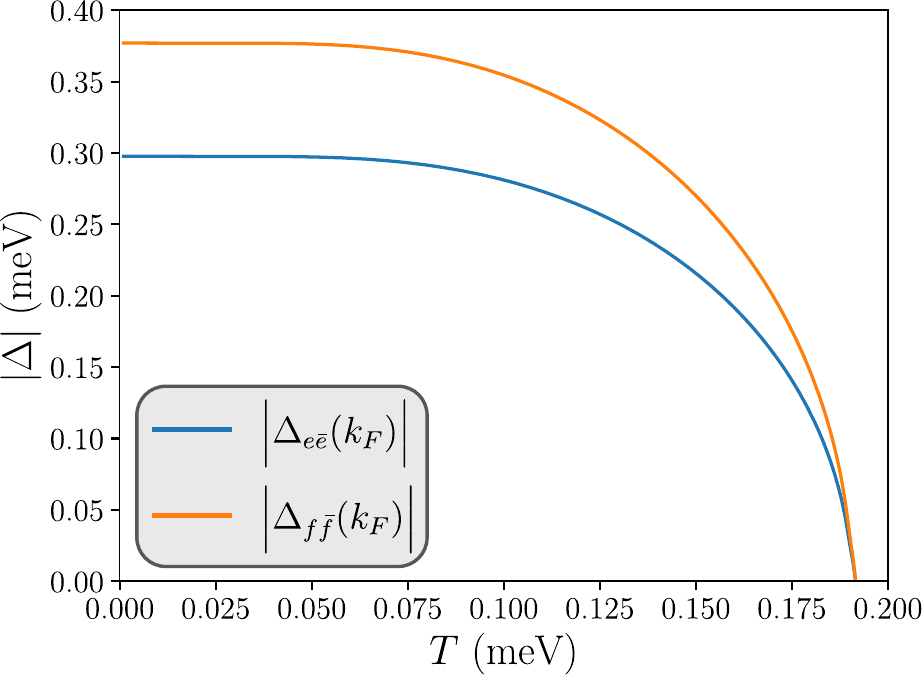}
\end{center}
\caption{s-wave superconducting gaps $|\Delta_{e\bar{e}}(k_F)|$ (blue line) and $|\Delta_{f\bar{f}}(k_F)|$ (orange line), evaluated at the corresponding Fermi surfaces, as functions of the temperature. The gap functions are obtained from solving the non-linear gap equation. We have used $U_1=0$, $U_2=2.32$ eV and $\omega_D=0.1$ eV, to reproduce the measured s-wave gap $|\Delta_s| \approx  0.3\; \text{meV}$ and $T_c\approx0.19$ meV$\approx2.5$ K at $h=0$ (Ref. \cite{Powell_2025}).}

\label{fig:4}
\end{figure}

We again find that the interaction contains an attractive piece acting between $e/{\bar e}$ and $f/{\bar f}$ components and a  repulsive piece acting within $e/{\bar e}$ and  $f/{\bar f}$ components. As in the case of the intra-band pairing,  the sign change leading to an attraction between $e/{\bar e}$ and $f/{\bar f}$ components  comes from the coherence factors from the diagonalization of the quadratic Hamiltonian.  
At the bare level, each component is of value $U$ as was the case in the p-wave channel. The degeneracy is again broken by the Kohn-Luttinger-type renormalization.  However, for the s-wave case, this renormalization acts against pairing as a repulsive intra-pocket interaction component gets larger than the attractive inter-pocket one (see Appendix \ref{AppendixB} for details). There is another, more conventional route towards attraction in the s-wave channel: the phonon-mediated attraction. In the standard Morel-Anderson picture \cite{Morel_1962}, the  repulsive $U$ is reduced at low frequencies by downward renormalization in the particle-particle channel, and if this downward renormalization is strong enough (which is the case when the Debye frequency is far smaller than the Fermi energy), the strength of the renormalized electron-electron interaction becomes smaller than electron-phonon attraction. For our system, this mechanism implies that at low-frequencies  a repulsive $U$ in (\ref{eq:22}) may change sign and become attractive. The solution of (\ref{eq:22}) then yields an  $s^\pm$ superconducting state below a finite $T_c$ (the signs of $\Delta_{e{\bar e}}$ and $\Delta_{f{\bar f}}$ are opposite). On a more careful examination we note that  even if the electron-phonon interaction is not strong enough to flip the sign of $U$, it still reduces  the intra-pocket interaction more than the intra-pocket one as the latter  involves a larger momentum transfer. If this effect is stronger than the Kohn-Luttinger  renormalization, we have $U_1 < U_2$, which is  the condition  for the s-wave instability in Eq. (\ref{eq:22}).  If $U_2$ remains positive, the superconducting order parameter has the same sign on the two pockets, if it changes sign,  the superconducting state is $s^\pm$.  This distinction is not essential to our analysis and for definiteness we keep $U_2 >0$ and set the values of $U_2$ and $U_1$ to match the experimental data for the s-wave gap. We solve the non-linear gap equation for s-wave pairing
\begin{equation}
\Delta_{\alpha\bar{\alpha}}\left(\textbf{k}\right)=-\sum_{\beta} \intop\frac{d^{2}\textbf{p}}{\left(2\pi\right)^{2}}\frac{U^{\text{inter}}_{\alpha\bar{\alpha};\bar{\beta}\beta}(\textbf{k},\textbf{p})\Delta_{\beta\bar{\beta}}\left(\textbf{p}\right)}{2E_{\beta\bar{\beta}}(\textbf{p})}\tanh\left(\frac{E_{\beta\bar{\beta}}(\textbf{p})}{2T}\right)
\label{eq:20}
\end{equation}
using  $U^{\text{inter}}_{e\bar{e};\bar{e}e} = U_1$ and $U^{\text{inter}}_{e\bar{e};\bar{f}f} =-U_2$ and show the results in Fig. \ref{fig:4}. We see that the $T_c$ and the gap values at $T=0$ are much larger for the s-wave case than for p-wave for comparable  $U_{1,2}$.  The reason for this is that $\tilde{L}_\pm$ and $L_\pm$ have different prefactors for $\log{1.13 \omega_D}{T}$:  the one in  $L_\pm$  has the  extra factor $\alpha^2k_{F\pm}^2/(\varepsilon^2+\alpha^2k_{F\pm}^2)$, see Eq. (\ref{eq:3.6}). This extra factor is not large -- it is  $0.845$ for $k_{F +}$ and $0.628$ for $k_{F -}$. Yet, at weak coupling  this extra factor  does reduce $T_c$ and  the gap at $T=0$  for p-wave pairing by roughly a factor of $4$, (see Figs. \ref{fig:3} and \ref{fig:4}).  Given this and also the very fact that the effect of electron-phonon interaction is generally  stronger than that of Kohn-Luttinger-type renormalization, we assume  that at zero field superconductivity is s-wave. The corresponding Bogoliubov dispersion has no nodes.

\vspace{-5mm}
\section{Finite magnetic field}
\label{secIV}

\subsection{Diagonalization of the non-interacting Hamiltonian}
\label{secIVA}
We now move to a finite magnetic field. The Hamiltonian is now given by
\begin{align}
\mathcal{H}_{0}= & \sum_{\textbf{k},s}\left(tk^{2}-\mu\right)a_{\textbf{k},s}^{\dagger}a_{\textbf{k},s}+\sum_{\textbf{k},s}\left(tk^{2}-\mu\right)b_{\textbf{k},s}^{\dagger}b_{\textbf{k},s}-\varepsilon\sum_{\textbf{k},s}\left(a_{\textbf{k},s}^{\dagger}b_{\textbf{k},s}+b_{\textbf{k},s}^{\dagger}a_{\textbf{k},s}\right)\nonumber \\
+ & \sum_{\textbf{k},s}\left[\left(-h+i\alpha ke^{i\varphi_{k}}\right)a_{\textbf{k},\downarrow}^{\dagger}a_{\textbf{k},\uparrow}-\left(h+i\alpha ke^{-i\varphi_{k}}\right)a_{\textbf{k},\uparrow}^{\dagger}a_{\textbf{k},\downarrow}\right]\nonumber \\
+ & \sum_{\textbf{k},s}\left[\left(-h+i\alpha ke^{-i\varphi_{k}}\right)b_{\textbf{k},\uparrow}^{\dagger}b_{\textbf{k},\downarrow}-\left(h+i\alpha ke^{i\varphi_{k}}\right)b_{\textbf{k},\downarrow}^{\dagger}b_{\textbf{k},\uparrow}\right]
\end{align}
The transformation that diagonalizes this Hamiltonian is 
\begin{align}
a_{\textbf{k},\uparrow}&=\frac{1}{\sqrt{2}}\left[\tilde z_{\textbf{k}}g_{\textbf{k}}+\tilde z_{\textbf{k}}j_{\textbf{k}}+\tilde\gamma_{\textbf{k}}\bar{g}_{\textbf{k}}+\tilde\gamma_{\textbf{k}}\bar{j}_{\textbf{k}}\right]\nonumber\\
a_{\textbf{k},\downarrow}&=\frac{1}{\sqrt{2}}\left[\tilde z_{\textbf{k}}e^{i\delta_{\textbf{k}}}g_{\textbf{k}}-\tilde z_{\textbf{k}}e^{i\delta_{\textbf{k}}}j_{\textbf{k}}-\tilde\gamma_{\textbf{k}}e^{-i\tilde{\delta}_{\textbf{k}}}\bar{g}_{\textbf{k}}+\tilde\gamma_{\textbf{k}}e^{-i\tilde{\delta}_{\textbf{k}}}\bar{j}_{\textbf{k}}\right]\nonumber\\
b_{\textbf{k},\uparrow}&=\frac{1}{\sqrt{2}}\left[-\tilde\gamma_{\textbf{k}}e^{i\delta_{\textbf{k}}}g_{\textbf{k}}+\tilde\gamma_{\textbf{k}}e^{i\delta_{\textbf{k}}}j_{\textbf{k}}-\tilde z_{\textbf{k}}e^{-i\tilde{\delta}_{\textbf{k}}}\bar{g}_{\textbf{k}}+\tilde z_{\textbf{k}}e^{-i\tilde{\delta}_{\textbf{k}}}\bar{j}_{\textbf{k}}\right]\nonumber\\
b_{\textbf{k},\downarrow}&=\frac{1}{\sqrt{2}}\left[-\tilde\gamma_{\textbf{k}}g_{\textbf{k}}-\tilde\gamma_{\textbf{k}}j_{\textbf{k}}+\tilde z_{\textbf{k}}\bar{g}_{\textbf{k}}+\tilde z_{\textbf{k}}\bar{j}_{\textbf{k}}\right]
\end{align}
where we have defined
\begin{align}
\begin{Bmatrix}\tilde\gamma_{\textbf{k}}\\
\tilde z_{\textbf{k}}
\end{Bmatrix}&=\frac{1}{\sqrt{2}}\left[1\pm\frac{\alpha k_y}{\sqrt{\varepsilon^{2}+\alpha^{2}k_y^{2}}}\right]^{1/2}\nonumber\\
\begin{Bmatrix}\delta_{\textbf{k}}\\
\tilde{\delta}_{\textbf{k}}
\end{Bmatrix}&=\text{Arg}\left[\sqrt{\varepsilon^{2}+\alpha^{2}k_{y}^{2}}\mp h+iak_{x}\right]
\end{align}
In the new basis, where $\mathcal{H}_{0}$ is diagonal, there are four bands which are now split and non circular.
\begin{align}
\mathcal{H}_{0}= & \sum_{\textbf{k}}\left[tk^{2}-\mu+\sqrt{\alpha^{2}k^{2}+\varepsilon^{2}+h^{2}-2h\sqrt{\alpha^{2}k_{y}^{2}+\varepsilon^{2}}}\right]g_{\textbf{k}}^{\dagger}g_{\textbf{k}}\nonumber \\
+ & \sum_{\textbf{k}}\left[tk^{2}-\mu+\sqrt{\alpha^{2}k^{2}+\varepsilon^{2}+h^{2}+2h\sqrt{\alpha^{2}k_{y}^{2}+\varepsilon^{2}}}\right]\bar{g}_{\textbf{k}}^{\dagger}\bar{g}_{\textbf{k}}\nonumber \\
+ & \sum_{\textbf{k}}\left[tk^{2}-\mu-\sqrt{\alpha^{2}k^{2}+\varepsilon^{2}+h^{2}-2h\sqrt{\alpha^{2}k_{y}^{2}+\varepsilon^{2}}}\right]j_{\textbf{k}}^{\dagger}j_{\textbf{k}}\nonumber \\
+ & \sum_{\textbf{k}}\left[tk^{2}-\mu-\sqrt{\alpha^{2}k^{2}+\varepsilon^{2}+h^{2}+2h\sqrt{\alpha^{2}k_{y}^{2}+\varepsilon^{2}}}\right]\bar{j}_{\textbf{k}}^{\dagger}\bar{j}_{\textbf{k}}
\end{align}

The fermions $g_{\textbf{k}}$, $\bar{g}_{\textbf{k}}$ and $j_{\textbf{k}}$, $\bar{j}_{\textbf{k}}$ are descendants of the fermions $e_{\textbf{k}}$, $\bar{e}_{\textbf{k}}$ and $f_{\textbf{k}}$, $\bar{f}_{\textbf{k}}$ respectively, but there is no one-to-one correspondence at $h=0$. The reason for this discrepancy, as already discussed, is that at $h=0$ any unitary transformation mixing $e_{\textbf{k}}$-$\bar{e}_{\textbf{k}}$ and $f_{\textbf{k}}$-$\bar{f}_{\textbf{k}}$ is still a transformation that diagonalizes the Hamiltonian, due to the bands being doubly degenerate. On the other hand the magnetic field lifts this degeneracy (Fig. \ref{fig:2b}) resulting in four distinct bands. In other words, the magnetic field selects the basis that will diagonalize the full Hamiltonian departing from the $e_{\textbf{k}}$, $\bar{e}_{\textbf{k}}$ and $f_{\textbf{k}}$, $\bar{f}_{\textbf{k}}$ fermions that were convenient to work with before. Even so, the two bases should be equivalent at zero field, because at zero field, the $e_{\textbf{k}}$, $\bar{e}_{\textbf{k}}$ and $f_{\textbf{k}}$, $\bar{f}_{\textbf{k}}$ fermions form two degenerate doublets, and the transformation to the $g_{\textbf{k}}$, $\bar{g}_{\textbf{k}}$ and $j_{\textbf{k}}$, $\bar{j}_{\textbf{k}}$ basis can be viewed as a momentum dependent pseudospin rotation within each doublet. To verify the equivalence of the two bases we define the transformation matrix that relates the two in the following way
\begin{align}
\psi_{0,\textbf{k}}^\dagger=A(h,\textbf{k})\psi_\textbf{k}^\dagger
\label{eq:25}
\end{align}
where $\psi_{0,\textbf{k}}^\dagger=\begin{pmatrix}e_{\textbf{k}}^\dagger & \bar{e}_{\textbf{k}}^\dagger & f_{\textbf{k}}^\dagger & \bar{f}_{\textbf{k}}^\dagger \end{pmatrix}^{T}$ and $\psi_{\textbf{k}}^\dagger=\begin{pmatrix}g_{\textbf{k}}^\dagger & \bar{g}_{\textbf{k}}^\dagger & j_{\textbf{k}}^\dagger & \bar{j}_{\textbf{k}}^\dagger \end{pmatrix}^{T}$. $A(h,\textbf{k})$ is a 4x4 matrix that relates the two different bases. At magnetic fields of the order of $h \lesssim 10^{-4}$ eV, $A(h,\textbf{k})$ is very weakly dependent on $h$ and the dependence on it can be safely neglected. The analytical expression for $A(\textbf{k})$ is rather complicated and therefore is not presented here. However, for analytical calculations, $A(\textbf{k})$ can be approximated as
\begin{equation}
A(\textbf{k})\approx\tfrac{1}{\sqrt{2}}\begin{pmatrix}
ie^{i\phi_k} & -ie^{i\phi_k} & 0 & 0 \\
-ie^{-i\phi_k} & -ie^{-i\phi_k} & 0 & 0 \\
0 & 0 & ie^{-i\phi_k} & ie^{-i\phi_k} \\
0 & 0 & ie^{i\phi_k} & -ie^{i\phi_k}
\end{pmatrix}
\label{eq:26}
\end{equation}

\subsection{Re-evaluation of superconductivity at h=0}
\label{secIVB}
We first re-evaluate the gap structure in the $g$-$j$ basis and at zero field. The analysis is more complex than in the $e$-$f$ basis (that is the reason why we chose this basis for the computations at $h=0$), yet the results are  easily  understandable. 
\vspace{-5mm}
\subsubsection{p-wave channel}
Using Eq. (\ref{eq:26}), the superconducting gap components of the p-wave channel can be expressed in the new basis as
\begin{align}
\Delta_{gg}(\textbf{k})&=\Delta_{\bar{g}\bar{g}}(\textbf{k})=\frac{e^{2i\varphi_k}\Delta_{ee}(\textbf{k})+e^{-2i\varphi_k}\Delta_{\bar{e}\bar{e}}(\textbf{k})}{2}=\Delta_{ee}\gamma_k z_k\cos\varphi_k\nonumber\\
\Delta_{g\bar{g}}(\textbf{k})&=\Delta_{\bar{g}g}(\textbf{k})=\frac{e^{-2i\varphi_k}\Delta_{\bar{e}\bar{e}}(\textbf{k})-e^{2i\varphi_k}\Delta_{ee}(\textbf{k})}{2}=-i\Delta_{ee}\gamma_k z_k\sin\varphi_k
\label{eq:27}
\end{align}
\begin{align}
\Delta_{jj}(\textbf{k})&=\Delta_{\bar{j}\bar{j}}(\textbf{k})=\frac{e^{-2i\varphi_k}\Delta_{ff}(\textbf{k})+e^{2i\varphi_k}\Delta_{\bar{f}\bar{f}}(\textbf{k})}{2}=\Delta_{ff}\gamma_k z_k\cos\varphi_k\nonumber\\
\Delta_{j\bar{j}}(\textbf{k})&=\Delta_{\bar{j}j}(\textbf{k})=\frac{e^{-2i\varphi_k}\Delta_{ff}(\textbf{k})-e^{2i\varphi_k}\Delta_{\bar{f}\bar{f}}(\textbf{k})}{2}=-i\Delta_{ff}\gamma_k z_k\sin\varphi_k
\label{eq:28}
\end{align}
We note here that in this new basis, there are both intraband, as well as interband components of equal magnitude. The pairing symmetry is still p-wave. More specifically, diagonal components have a $p_x$ symmetry whereas off-diagonal components have a $p_y$ symmetry. The equivalence of the two bases can be immediately understood from the BdG excitation spectrum which should remain invariant under the basis rotation. Indeed, the excitation spectrum of the BdG Hamiltonian consists of two doubly degenerate bands which are given by
\begin{align}
E_{1}(\textbf{k})&=\sqrt{\left(\varepsilon_{+}(k)\right)^2+|\Delta_{gg}(\textbf{k})|^2+|\Delta_{g{\bar{g}}}(\textbf{k})|^2}=\sqrt{\left(\varepsilon_{+}(k)\right)^2+\Delta_{ee}^2\gamma_k^2z_k^2}\nonumber\\
E_{2}(\textbf{k})&=\sqrt{\left(\varepsilon_{-}(k)\right)^2+|\Delta_{jj}(\textbf{k})|^2+|\Delta_{j\bar{j}}(\textbf{k})|^2}=\sqrt{\left(\varepsilon_{-}(k)\right)^2+\Delta_{ff}^2\gamma_k^2z_k^2}
\label{eq:29}
\end{align}
We see that the angle dependence drops out from Eq. (\ref{eq:29}), as was the case in the $e$, $f$ fermion basis where $|\Delta(\textbf{k})|$ was independent on $\varphi_k$.
\subsubsection{s-wave channel}
The situation in the s-wave channel is relatively simpler. In this case the pairing remains purely interband even in the new basis. The non-zero superconducting gaps in this basis are related to the old ones in the following way:
\begin{align}
\Delta_{g\bar{g}}(\textbf{k})&=-\Delta_{e\bar{e}}(\textbf{k})\nonumber\\
\Delta_{j\bar{j}}(\textbf{k})&=-\Delta_{f\bar{f}}(\textbf{k})
\label{eq:30}
\end{align}
This is nothing other than multiplying the gap by a global phase which plays no role in the analysis of this channel at finite field.

\subsection{Superconductivity at finite magnetic field}
\label{secIVC}
To analyze superconductivity at a finite magnetic field for both channels, 
we need the pairing interaction in the $g$-$j$ basis. To get it, we first redefine the pairing interaction Hamiltonian in the $e$-$f$  basis as 

\begin{equation}
\mathcal{H}_{int}=\sum_{\textbf{k},\textbf{p}}U^0_{\alpha\beta;\gamma\delta}(\textbf{k},\textbf{p})[\psi_{0,\textbf{k}}^\dagger]_\alpha[\psi_{0,-\textbf{k}}^\dagger]_\beta[\psi_{0,-\textbf{p}}]_\gamma[\psi_{0,\textbf{p}}]_\delta
\end{equation}
where we remind that $\psi_{0,\textbf{k}}^\dagger=\begin{pmatrix}e_{\textbf{k}}^\dagger & \bar{e}_{\textbf{k}}^\dagger & f_{\textbf{k}}^\dagger & \bar{f}_{\textbf{k}}^\dagger \end{pmatrix}^{T}$. In the previous equation, each index is summed over all four fermionic flavors. This allows one to immediately see that in the new basis the interaction transforms in the following way:
\begin{equation}
U_{\alpha\beta;\gamma\delta}(\textbf{k},\textbf{p})=\sum_{\alpha'\beta'\gamma'\delta'}A_{\alpha'\alpha}(\textbf{k})A_{\beta'\beta}(-\textbf{k})A^*_{\gamma'\gamma}(-\textbf{p})A^*_{\delta'\delta}(\textbf{p})U^0_{\alpha'\beta';\gamma'\delta'}(\textbf{k},\textbf{p})
\label{eq:38}
\end{equation}
In the above notation, $A_{\alpha'\alpha}(\textbf{k})$ corresponds to the matrix elements of $A(\textbf{k})$ from Eq. (\ref{eq:25}). Primed indices run over $e$, $\bar{e}$, $f$, $\bar{f}$, while unprimed indices can take the values $g$, $\bar{g}$, $j$, $\bar{j}$. The system of linearized gap equations at a finite magnetic field reduces to
\begin{align}
\Delta_{\alpha\beta}\left(\textbf{k}\right)=-\sum_{\gamma\delta} \intop\frac{d^{2}\textbf{p}}{\left(2\pi\right)^{2}}\frac{U_{\alpha\beta;\gamma\delta}(\textbf{k},\textbf{p})\Delta_{\delta\gamma}\left(\textbf{p}\right)}{\varepsilon_{\gamma}(\textbf{p},h)+\varepsilon_{\delta}(\textbf{p},h)}\left[\tanh\left(\frac{\varepsilon_{\gamma}(\textbf{p},h)}{2T_c}\right)+\tanh\left(\frac{\varepsilon_{\delta}(\textbf{p},h)}{2T_c}\right)\right]
\label{eq:39}
\end{align}
where the pairing matrix is defined as
\begin{equation}
\hat{\Delta}(\textbf{k})=
\begin{pmatrix}
\Delta_{gg}(\textbf{k}) & \Delta_{g\bar{g}}(\textbf{k}) & 0 & 0 \\
\Delta_{\bar{g}g}(\textbf{k}) & \Delta_{\bar{g}\bar{g}}(\textbf{k}) & 0 & 0 \\
0 & 0 & \Delta_{jj}(\textbf{k}) & \Delta_{j\bar{j}}(\textbf{k}) \\
0 & 0 & \Delta_{\bar{j}j}(\textbf{k}) & \Delta_{\bar{j}\bar{j}}(\textbf{k})
\end{pmatrix}
\end{equation}
with $\Delta_{\bar{g}g}(\textbf{k})=-\Delta_{g\bar{g}}(-\textbf{k})$ and $\Delta_{\bar{j}j}(\textbf{k})=-\Delta_{j\bar{j}}(-\textbf{k})$.

\subsubsection{p-wave channel}
\begin{figure}[H]
\centering
\begin{minipage}{.32\textwidth}
    \centering
    \includegraphics[width=\linewidth,height=4cm]{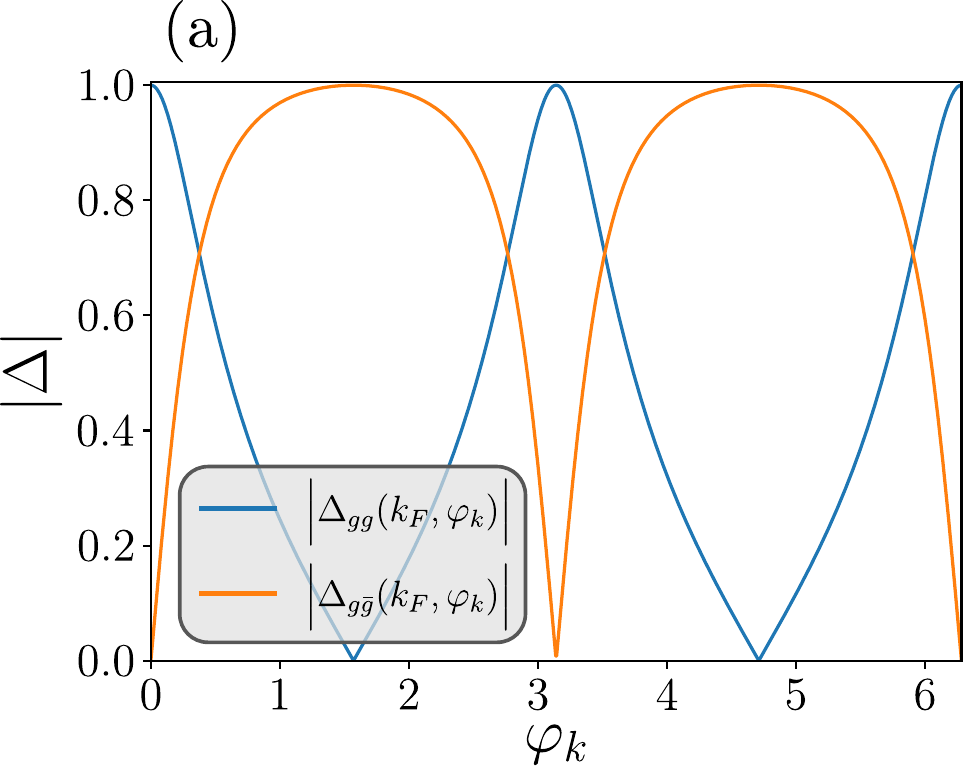}
    \phantomsubcaption\label{fig:5a}
\end{minipage}
\hfill
\begin{minipage}{.32\textwidth}
    \centering
    \includegraphics[width=\linewidth,height=4cm]{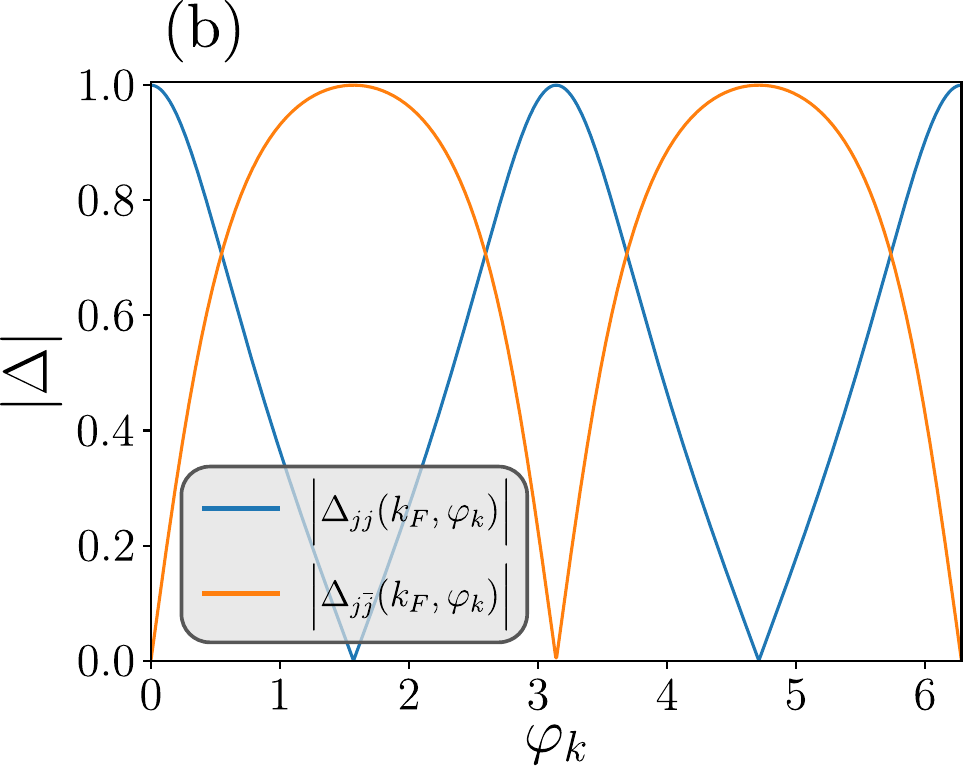}
    \phantomsubcaption\label{fig:5b}
\end{minipage}
\hfill
\begin{minipage}{.32\textwidth}
    \centering
    \includegraphics[width=\linewidth,height=4cm]{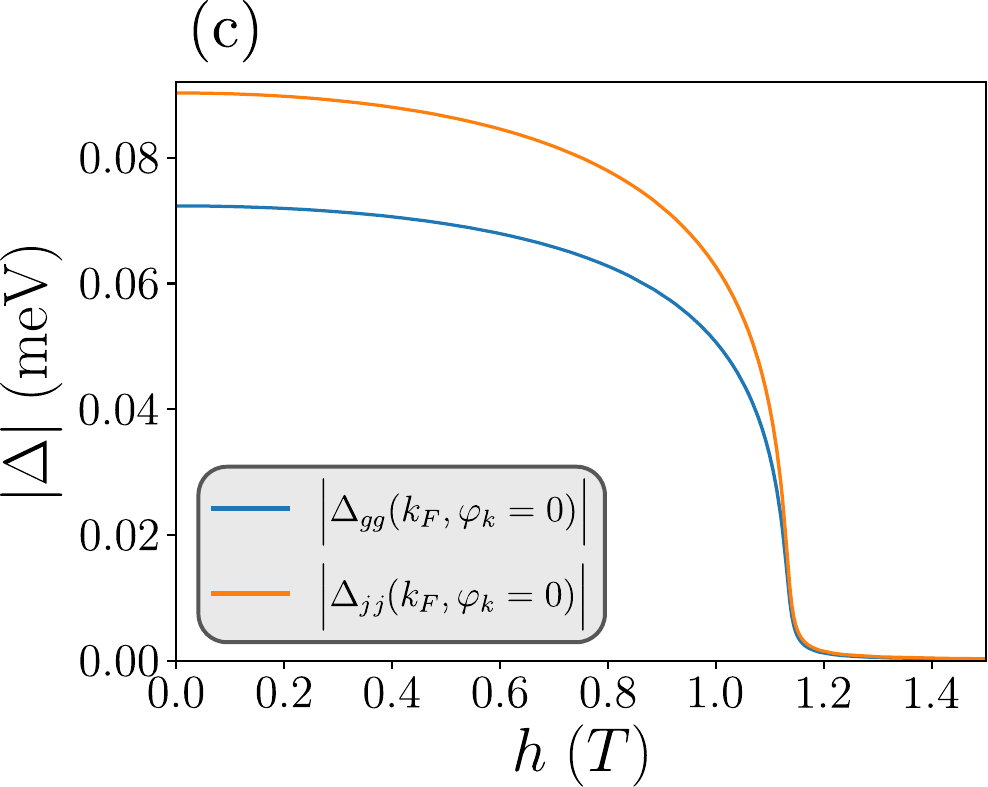}
    \phantomsubcaption\label{fig:5c}
\end{minipage}
\caption{(a)-(b) Polar angle dependence of the components of the gap functions on the large and small Fermi pockets respectively, evaluated at the Fermi surface. The blue lines correspond to the intraband components $|\Delta_{gg}(\varphi_k)|$, $|\Delta_{jj}(\varphi_k)|$, while the orange lines correspond to the interband components $|\Delta_{g\bar{g}}(\varphi_k)|$, $|\Delta_{j\bar{j}}(\varphi_k)|$. (c) Magnetic field dependence of the gap functions $\Delta_{gg}(\varphi_k=0)$ (blue line) and $\Delta_{jj}(\varphi_k=0)$ (orange line) evaluated at the Fermi surface. We have used $U_1=0$, $U_2=2.6$ eV and $\omega_D=0.1$ eV.}
\label{fig:5}
\end{figure}

Consider first the p-wave channel. Since we want to study the evolution of the gap under a finite magnetic field at temperatures $T\ll T_c$, we need to solve the non-linear gap equation. Applying Eq. (\ref{eq:38}) to transform $\mathcal{H}_{int}^{\text{intra}}$ into the $g$-$j$ basis, we find that all components of the interaction tensor become nonzero. As a consequence, both the intraband and interband components of the matrix gap function  become nonzero. This greatly complicates the analysis of the nonlinear gap equation when a magnetic field lifts the band degeneracy. Below we adopt an iterative approach in which we express the matrix gap equation as 
\begin{equation}
\Delta_{\alpha\beta}(\textbf{k})=-2\sum_{\textbf{p},\gamma\delta}U_{\alpha\beta;\gamma\delta}(\textbf{k},\textbf{p})\langle[\psi_{-\textbf{p}}]_\gamma[\psi_{\textbf{p}}]_\delta\rangle
\label{eq:41}
\end{equation}
and solve it iteratively. Namely,  we evaluate the $\langle \psi \psi \rangle$ expectation values at each iteration by diagonalizing the Bogoliubov–de Gennes (BdG) Hamiltonian,
\begin{equation}
\mathcal{H}_{BdG}=\frac{1}{2}\sum_{\textbf{k}}\begin{pmatrix}
\psi_\textbf{k}^\dagger & \psi_{-\textbf{k}}
\end{pmatrix}
\begin{pmatrix}
\xi(\textbf{k}) & -\hat{\Delta}(\textbf{k}) \\
-\hat{\Delta}(\textbf{k})^* & -\xi(-\textbf{k})
\end{pmatrix}
\begin{pmatrix}
\psi_\textbf{k}  \\
\psi_{-\textbf{k}}^\dagger 
\end{pmatrix}
\label{eq:42}
\end{equation}
where $\xi_{\alpha\beta}(\textbf{k})=\varepsilon_\alpha(\textbf{k})\delta_{\alpha\beta}$, and $\hat{\Delta}(\textbf{k})$ taken from the previous iteration. We use an initial guess of the form
\begin{equation}
\Delta_{\alpha\beta}(\textbf{k})=\Delta_{\alpha\beta}\gamma_kz_k\sum_{\gamma'}A_{\gamma'\alpha}(\textbf{k})A_{\gamma'\beta}(
-\textbf{k})\exp\left[-i\phi_k  c_{\gamma'}\right]
\end{equation}
where $\Delta_{\alpha\beta}$ is a magnetic field dependent but $k$-independent factor and $c_{\gamma'} = 1$ for $e$, $\bar{f}$  and $-1$ for $\bar{e}$, $f$. This Ansatz is chosen such that it reproduces the angular dependence obtained from solving the linearized gap equation (\ref{eq:39}), leaving the coefficients $\Delta_{\alpha\beta}$ as the only unknown parameters. We numerically carry out this iteration process, using the exact expression for $A(k)$, and show our results in Fig. \ref{fig:5}. We set $T=0.3\,$K -the same as in \cite{Powell_2025}. 

\begin{figure}[H]
\centering
\begin{minipage}{.46\textwidth}
    \centering
    \includegraphics[width=\linewidth,height=4cm]{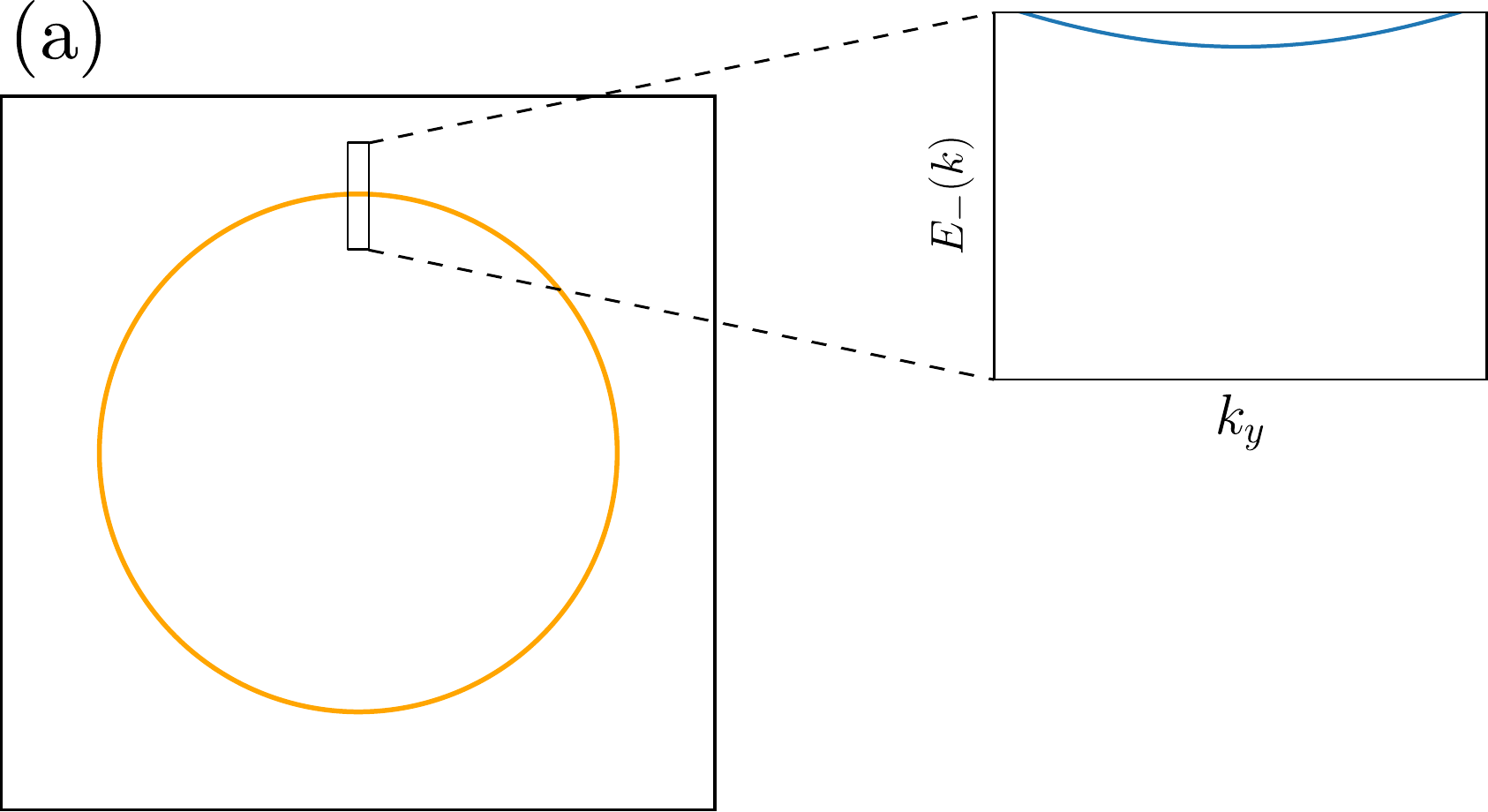}
    \phantomsubcaption\label{fig:6a}
\end{minipage}
\hfill
\begin{minipage}{.46\textwidth}
    \centering
    \includegraphics[width=\linewidth,height=4cm]{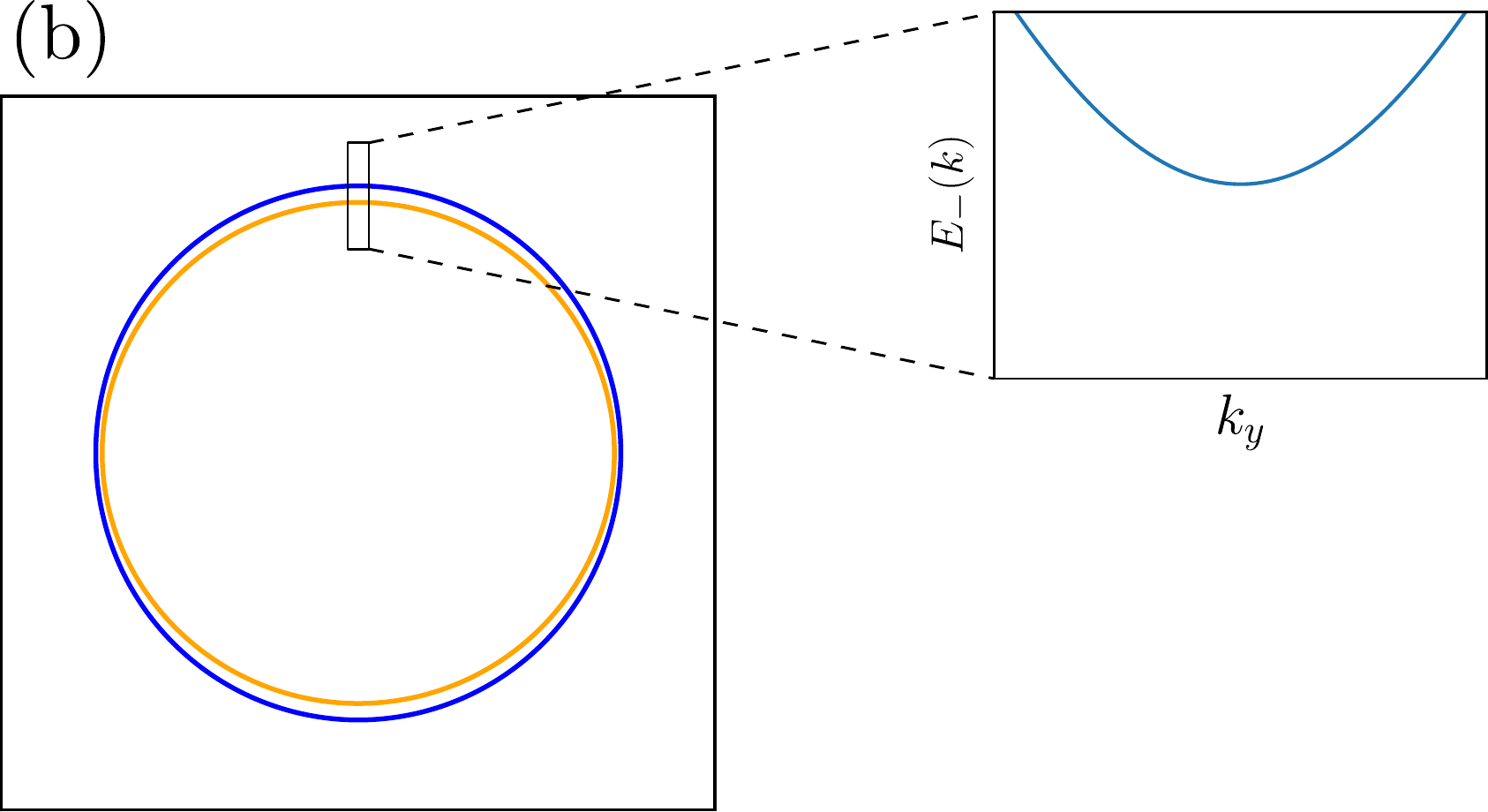}
    \phantomsubcaption\label{fig:6b}
\end{minipage}
\vspace{-1cm}
\begin{minipage}{.46\textwidth}
    \centering
    \includegraphics[width=\linewidth,height=4cm]{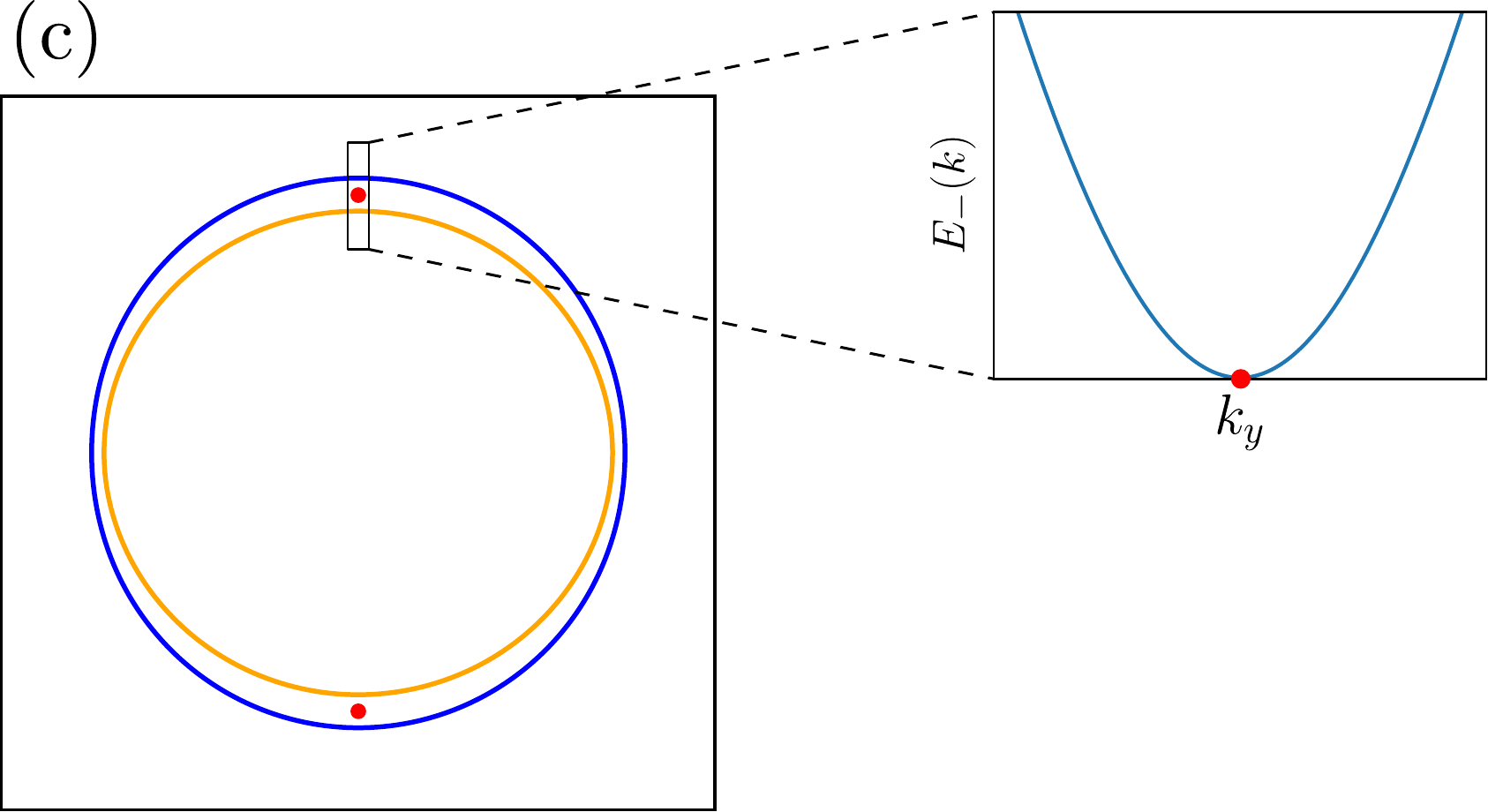}
    \phantomsubcaption\label{fig:6c}
\end{minipage}
\hfill
\begin{minipage}{.46\textwidth}
    \centering
    \includegraphics[width=\linewidth,height=4cm]{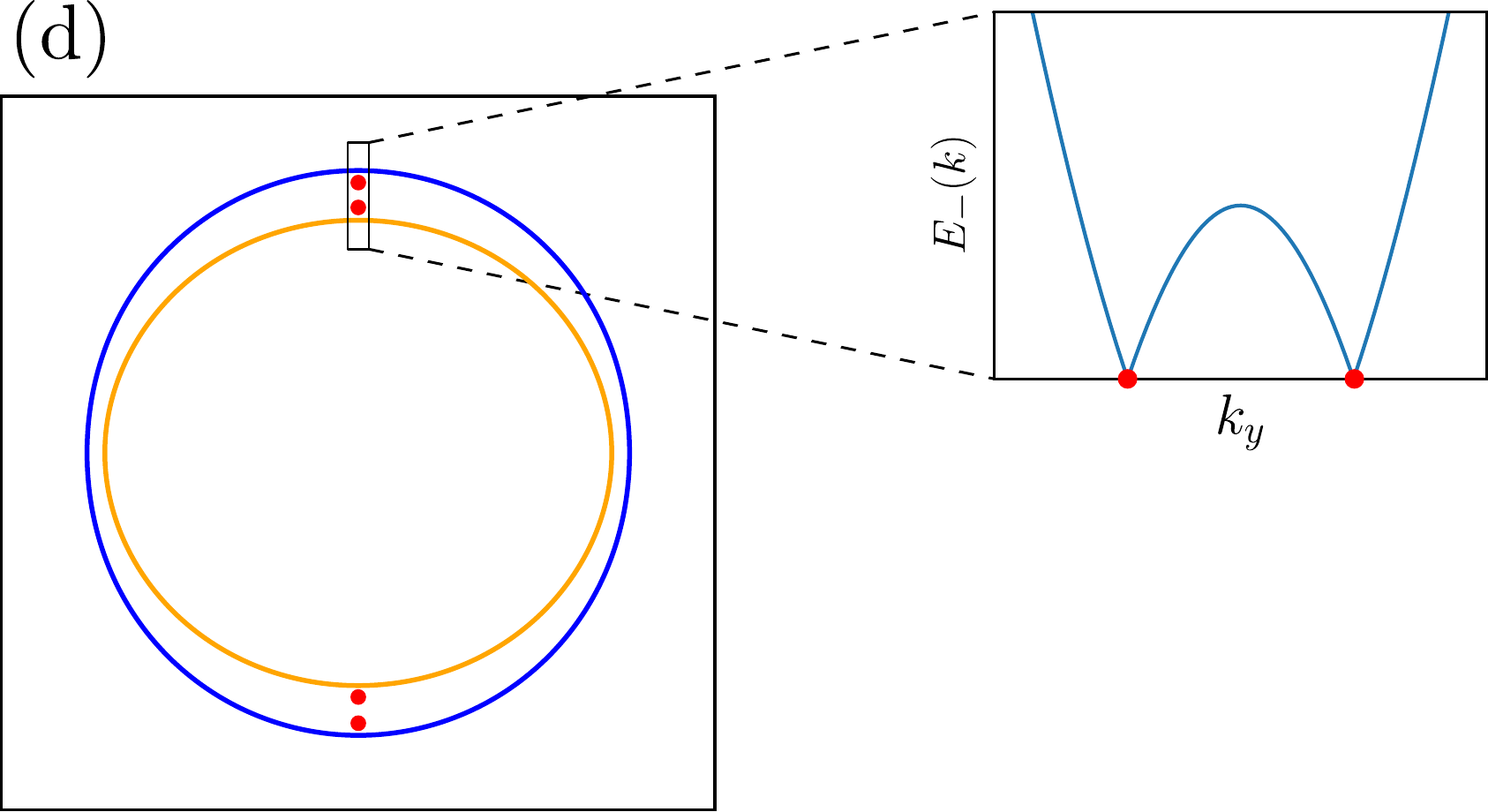}
    \phantomsubcaption\label{fig:6d}
\end{minipage}
\caption{Evolution of the superconducting excitation spectrum in the p-wave channel as a function of magnetic field. The main panels correspond to the larger Fermi pockets of the normal state, while the insets present the BdG Hamiltonian lower band near $\varphi_k=\pi/2$ for (a) $h=0$, (b) $h<\Delta_{gg}$, (c) $h=\Delta_{gg}$ and (d) $h>\Delta_{gg}$.}
\label{fig:6}
\end{figure}

We verified that the gap functions satisfy the relations $\Delta_{gg}({\bf k})=\Delta_{\bar{g}\bar{g}}({\bf k})$, $\Delta_{jj}({\bf k})=\Delta_{\bar{j}\bar{j}}({\bf k})$ and that the overall momentum-independent prefactors in $\Delta_{gg}({\bf k})$ and $\Delta_{g\bar{g}}({\bf k})$ as well as in $\Delta_{jj}({\bf k})$ and $\Delta_{j\bar{j}}({\bf k})$ are  also the  same. This implies that the system of gap equations is reduced to two  parameters $\Delta_{gg}(h)$ and $\Delta_{jj}(h)$ that we solve for. As shown in Figs. \ref{fig:5a} and \ref{fig:5b}, the gap functions exhibit an approximate angular dependence of the form 
$|\Delta_{gg}({\bf k})|\sim|\cos\varphi_k|$, $|\Delta_{jj}({\bf k})|\sim|\cos\varphi_k|$ and $|\Delta_{g\bar{g}}({\bf k})|\sim|\sin\varphi_k|$, $|\Delta_{j\bar{j}}({\bf k})|\sim|\sin\varphi_k|$, matching Eqs. (\ref{eq:27})-(\ref{eq:28}). In Fig. \ref{fig:5c} we present the evolution of the gap amplitude as a function of the applied magnetic field. We see that the gap smoothly decreases and vanishes at a critical magnetic field $h_c\approx1.3\;\text{T}$. The BdG excitation spectrum at a finite $h$  consists of  two sets of $E_{\pm}  ({\bf k})$ - one for $g $ and ${\bar g}$  fermions, another for $j $ and ${\bar j}$. They display identical behavior. For brevity we present the result  for $g $ and ${\bar g}$ fermions: 

\begin{equation}
E_{\pm}^2({\bf k})=\left(\sqrt{\xi_{+}^2({\bf k})+\Delta_{gg}^2\sin^2\varphi_k}\pm\Big|\xi_{-}({\bf k})\Big|\right)^2+\Delta_{gg}^2\cos^2\varphi_k
\end{equation}
where
\begin{equation}
\xi_{\pm}({\bf k})=\frac{\varepsilon_g({\bf k})\pm\varepsilon_{\bar{g}}({\bf k})}{2}
\end{equation}

At $h=0$ ($\xi_-=0$) the excitation spectrum is fully gapped and exhibits no dependence on the polar angle. However, at magnetic fields $h\geq \Delta_{gg}(h)$ the lower band develops nodes along the $k_y$ axis. This behavior is illustrated in Figs. \ref{fig:6a}-\ref{fig:6d} where it is evident that the band minimum eventually touches zero at a magnetic field $h=\Delta_{gg}$ at two ${\bf k}$-values along the $y$-axis: $k_y=\pm k_F(h=0)$. For  the parameters that we used this happens at  $h\gtrsim 0.9$ T. At a larger $h$, each zero splits into two nodal Dirac points that lie between where the normal state Fermi surfaces would be in the absence of superconductivity. This effect has been discussed before in \cite{Khodas_2012,Chubukov_2016} for a d-wave state, where the authors found that the nodes develop along the diagonal directions $\cos{2\varphi_k}=0$. Here, they develop along two directions specified by $\cos{\varphi_k}=0$. We emphasize that this is not a Bogoliubov Fermi surface \cite{Agterberg_2017,Brydon_2018}, since the zero-energy excitations remain point-like rather than forming extended surfaces in momentum space. 

\subsubsection{s-wave channel}

\begin{figure}[H]
\begin{center}
\includegraphics[width=.6\textwidth]{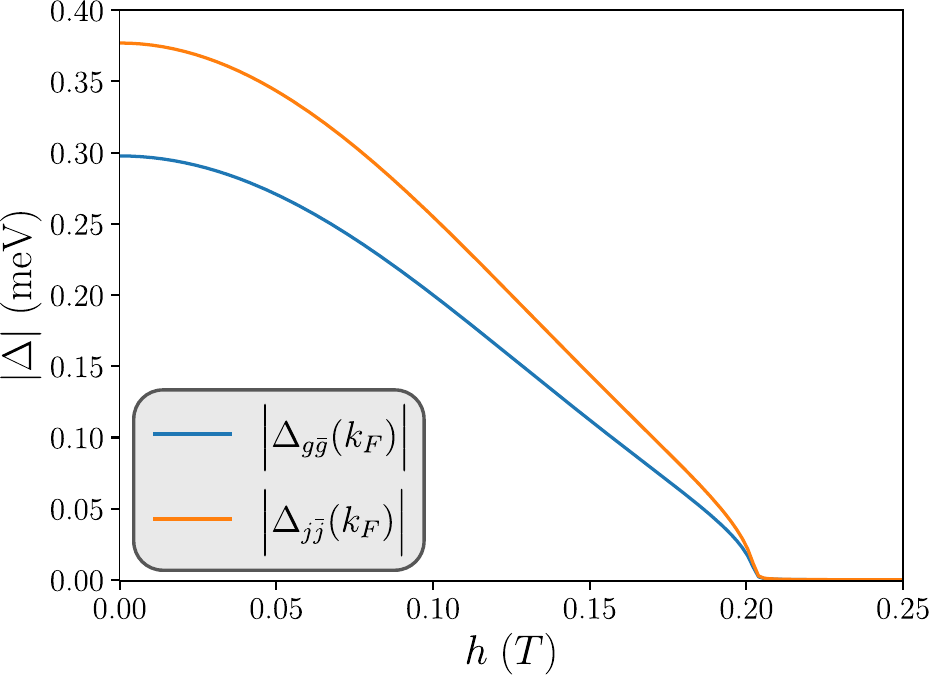}
\end{center}
\caption{Magnetic field dependence of the pairing matrix elements (gap functions) $\Delta_{g\bar{g}}(k_F)$ (blue line) and $\Delta_{j\bar{j}}(k_F)$ (orange line) evaluated at the Fermi surface. We have used $U_1=0$, $U_2=2.32$ eV as initial values for the couplings and $\omega_D=0.1$ eV. The couplings are quadratically varied to the final values $U_1=0.2$ eV, $U_2=2.12$ eV in a range of $0.2$ T.}
\label{fig:7}
\end{figure}

We next analyze the s-wave pairing gap in the presence of a magnetic field. Applying Eq. (\ref{eq:38}) to transform $\mathcal{H}_{int}^{\text{inter}}$ into the new basis, we find that it remains invariant to good approximation under the basis rotation, as Eq. (\ref{eq:30}) suggests. We see that even in the $g$ and $j$ fermion basis the pairing in this channel remains purely interband. The non-linear gap equation for the off-diagonal elements of the pairing matrix $\Delta_{\alpha\beta}(\textbf{k})$ $(\alpha\neq\beta)$ at a finite magnetic field can now be written in the following way:
\begin{equation}
\Delta_{\alpha\beta}\left(\textbf{k}\right)=-\sum_{\gamma\neq\delta} \intop\frac{d^{2}\textbf{p}}{\left(2\pi\right)^{2}}\frac{U_{\alpha\beta;\gamma\delta}(\textbf{k},\textbf{p})\Delta_{\delta\gamma}\left(\textbf{p}\right)}{2\sqrt{\left(\varepsilon^+_{\gamma\delta}(\textbf{p})\right)^2+\Big|\Delta_{\gamma\delta}(\textbf{p})\Big|^2}}\left[\tanh\left(\frac{E^+_{\gamma\delta}(\textbf{p})}{2T}\right)+\tanh\left(\frac{E^-_{\gamma\delta}(\textbf{p})}{2T}\right)\right]
\label{eq:47}
\end{equation}
where $\varepsilon^\pm_{\alpha\beta}(\textbf{k}) \equiv \frac{\varepsilon_\alpha(\textbf{k})\pm\varepsilon_\beta(\textbf{k})}{2}$ and $E^\pm_{\alpha\beta}(\textbf{k}) \equiv \sqrt{(\varepsilon^+_{\alpha\beta})^2+\big|\Delta_{\alpha\beta}(\textbf{k})\big|^2}\pm\varepsilon^-_{\alpha\beta}$. Importantly this channel remains fully gapped at all fields. The numerical solutions of Eq. (\ref{eq:47}) for various values of the magnetic field are shown in Fig. \ref{fig:7}. 

We phenomenologically introduce a magnetic field dependent renormalization of the interaction in this channel, in order to reproduce the experimentally observed difference between the zero field gap scale and the transition field. Experimentally, the low-field state has a gap scale $\Delta_s \simeq 0.3\; \text{meV}$, while the high-field state has a substantially smaller gap scale $\Delta_p \simeq 0.1\; \text{meV}$, and the transition between the two occurs already at $h \simeq 0.2\; \text{T}$. For the transition to occur near this field, the s-wave state must lose its energetic advantage over the p-wave state; in practice, this happens when the s-wave gap has been reduced to a value smaller than the p-wave gap. This behavior cannot be obtained with field-independent couplings $U_1$ and $U_2$. With constant interactions, the mean-field gap equation gives a critical field controlled by the zero-field s-wave gap scale itself. For our case of $\Delta_s(0)\simeq 0.3\; \text{meV}$, this would give a critical field $h_c\simeq 0.9$ T, much larger than the experimentally observed transition field. To match the data, we allow the effective difference $U_2-U_1$ in the s-wave channel to decrease with a magnetic field. Proceeding with this assumption, we show in Fig. \ref{fig:7} that we can reproduce the experimental observation, that s-wave superconductivity is destroyed at a smaller critical field $h_c\approx 0.2\;\text{T}$. A reduction of the s-wave coupling with a field should be viewed as a phenomenological way of incorporating additional pair-breaking effects, most likely orbital in origin, which are not included explicitly in the purely Zeeman mean-field treatment, but are needed to account for the rapid suppression of s-wave superconductivity. We note in this regard that orbital effects are likely not negligible in $\beta$-PdBi$_2$, despite the layered structure of the material, as can be inferred from the relatively small anisotropy of the experimentally reported values for the in-plane and out of plane superconducting coherence lengths $\xi_{ab}\simeq22$ nm and $\xi_{c}\simeq17$ nm respectively \cite{Powell_2025}. Combining this with the reported value of the out of plane critical field $B_{c2,\text{orb}}^{\perp}\simeq 1$ T (Ref. \cite{Powell_2025}), and assuming that it is predominantly orbital in origin, we get a lower bound for the in-plane orbital critical field $B_{c2,\text{orb}}^{||}\simeq 1.3$ T. This is larger than the Zeeman critical field of approximately $B_{c2,\text{Zeeman}}^{||}\simeq 0.9$ T, but not by much. In view of this, our Zeeman-only treatment should then be regarded as an approximate treatment that captures the physics of the field-induced  transformation from s-wave to p-wave superconductivity, but to get the values of the critical fields one likely has to include orbital effects. 

\subsubsection{Free energies in s-wave and p-wave channels}

\begin{figure}[H]
\begin{center}
\includegraphics[width=.6\textwidth]{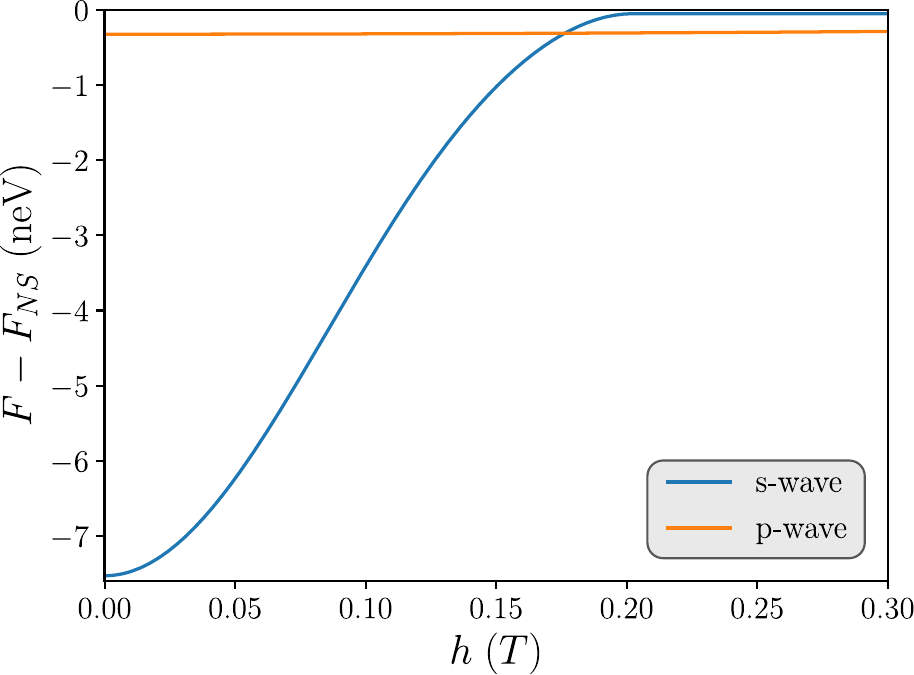}
\end{center}
\caption{Free energy difference between the superconducting and normal states for the two channels.}
\label{fig:8}
\end{figure}

In a superconducting state the free energy is given by
\begin{align}
F&=\frac{1}{2}\sum_{\textbf{k},\alpha\beta} \Delta_{\alpha\beta}(\textbf{k})\langle[\psi^\dagger_{\textbf{k}}]_\alpha[\psi^\dagger_{-\textbf{k}}]_\beta\rangle
+\frac{1}{2}\sum_{\textbf{k},\alpha}\Big[\varepsilon_\alpha(\textbf{k})-E_{\alpha}(\textbf{k})\Big]\nonumber\\
&-T\sum_{\textbf{k},\alpha}\log\Big[1+e^{-\frac{E_{\alpha}(\textbf{k})}{T}}\Big]+\mu N
\end{align}
where $E_{\alpha}(\textbf{k})$ are the four positive bands of the BdG Hamiltonian. We computed $F$ numerically in both s-wave and p-wave channels and show the results in Fig. \ref{fig:8}. As one would expect, we find that at zero and at small magnetic fields the free energy is smaller in the the p-wave channel, but at larger fields the Free energy of an s-wave superconductor becomes smaller.

\section{Differential Tunneling Conductance}
\label{secV}
To further compare our results with the data in \cite{Powell_2025}, we calculate the differential tunneling conductance  $G=\frac{dI}{dV}$. We  follow earlier works \cite{Tersoff_1985,Tinkham_2004} and relate $\frac{dI}{dV}$ to the weighted integral of the fermionic density of states:
\begin{equation}
G=\frac{dI}{dV}=\frac{G_0}{\bar{N}_{F}}\intop_{-\infty}^\infty N_{SC}(\omega)\frac{\partial f(\omega+eV_b)}{\partial(eV_b)}d\omega
\end{equation}
where $N_{SC}(\omega)$ is the total electronic density of states in the superconducting state, $f(\omega)$ is the Fermi-Dirac distribution, $G_0$ is the normal state conductance and $\bar{N}_{F}=2(N_{F+}+N_{F-})$ is the total density of states in the normal state: 
\begin{figure}[H]
\begin{center}
\includegraphics[width=.7\textwidth]{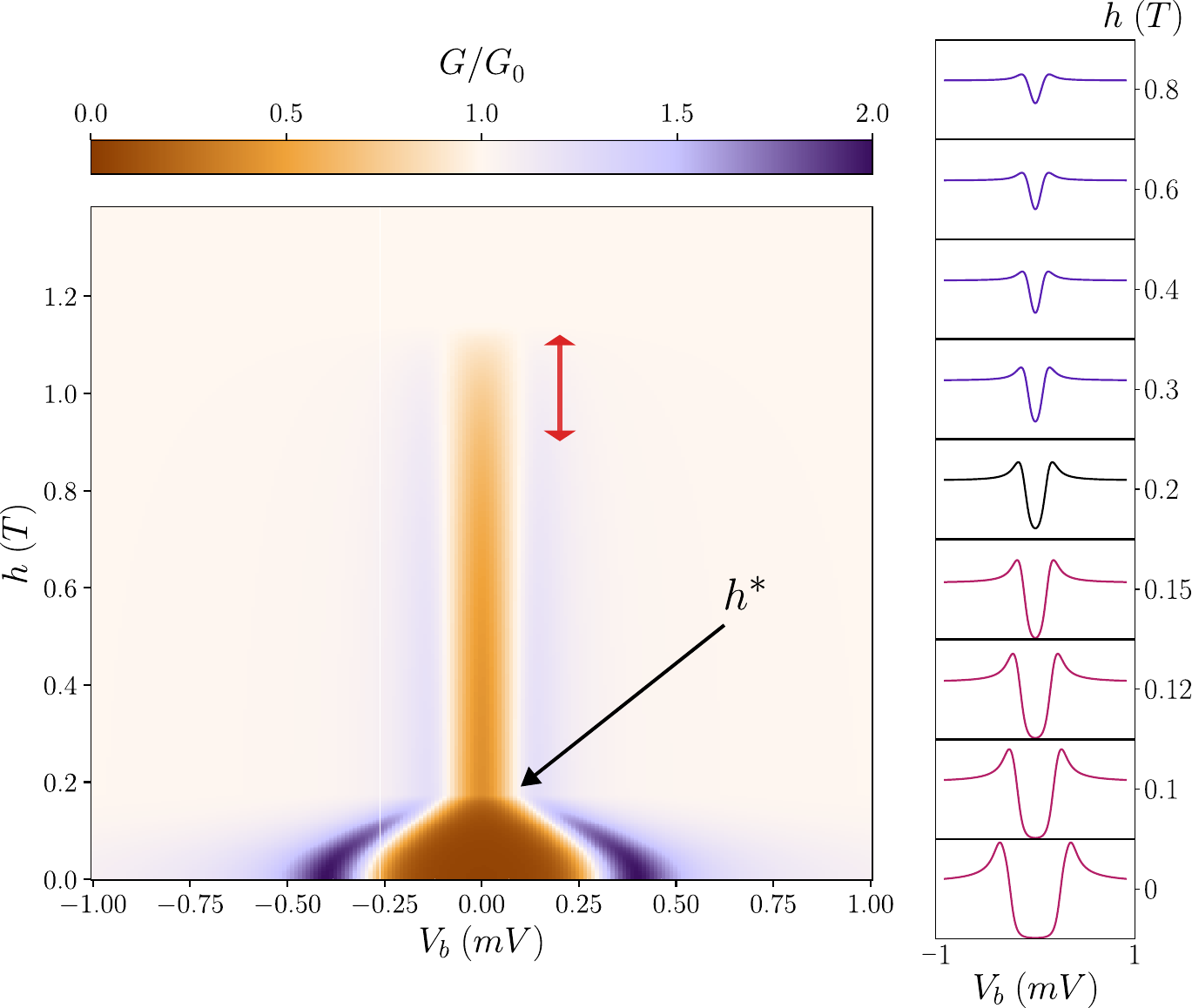}
\end{center}
\caption{Left: heatmap of the normalized differential tunneling conductance for an in plane magnetic field. The black arrow indicates the transition between s-wave and p-wave superconducting orders. The red double arrow marks the crossover to the nodal regime within the p-wave state. Right: $dI/dV$ at  selected values of the magnetic field. We clearly see the transformation from a U-shaped to a V-shaped  spectrum at fields around $h^*$.}
\label{fig:9}
\end{figure}

We use our numerical solutions for the gap functions and evaluate the differential conductance at different values of the magnetic field. We show the results in Fig. \ref{fig:9}. As expected, the first-order transition between the two superconducting channels manifests itself as a sharp feature in the differential tunneling conductance heatmap, marked by the black arrow. This behavior is fully consistent with the kink reported in \cite{Powell_2025}. Moreover, we observe a clear evolution of the conductance profile from a U-shaped form at fields smaller than $h^*$ to a V-shaped form at higher fields. In the latter field range, we also find an increasing zero bias conductance. We note that the magnetic field range over which the spectrum is V-shaped, is broader than the regime in which the Bogoliubov spectrum is actually nodal. We verified that this broadening is a finite temperature effect (we computed $dI/dV$ at a finite temperature $T=0.3$ K, the same as in \cite{Powell_2025}). Namely, when the minimal excitation energy becomes comparable to $T$, excitations behave as if they were gapless, and the conductance displays a V-shaped form even before the true nodal regime is reached.

This last observation provides a way for an experimental verification of the key aspect of our scenario that the transition to the p-wave state does not coincide with the crossover to a V-shaped tunneling spectrum. In a more  conventional scenario of a transition to a nodal p-wave, the V-shaped tunneling spectrum emerges right at the transition. We emphasize that in our theory the separation between the transition to p-wave and the emergence of a V-shaped tunneling spectrum is robust against variations of the couplings $U_1$ and $U_2$, provided that the condition $U_2>U_1$ is satisfied. Another interesting possibility is a possible emergence of an FFLO s-wave state before s-wave superconductivity yields to p-wave. In general, an s-wave FFLO state becomes energetically favorable relative to a uniform s-wave at sufficiently large magnetic fields, when the s-wave order parameter is strongly suppressed. In our case, a p-wave state becomes energetically favorable, while the s-wave order parameter is still finite.  In this situation, an FFLO state is still possible near the transition to p-wave, but by no means guaranteed.  

\section{Conclusions}
\label{secVI}
In this work, we have presented a microscopic theory for the magnetic field induced transition from nodeless to nodal superconductivity in $\beta$-PdBi$_2$ based on a competition between two pairing channels with different sensitivity to Zeeman-induced band splitting. At low magnetic fields the superconducting state is s-wave and excitations are fully gapped. An in-plane magnetic field rapidly suppresses this channel since pairing no longer occurs on the Fermi surface. A second p-wave channel then wins at higher fields. In this channel intra- and interband pairing components are comparable and therefore the pairing is more robust under an external magnetic field.

We have shown that in the high-field phase, the Bogoliubov spectrum eventually becomes gapless, with nodes developing in between the normal state magnetic field split Fermi surfaces. This provides a natural route to nodal superconductivity without invoking additional symmetry breaking and is consistent with the tunneling spectroscopy observations in $\beta$-PdBi$_2$. More broadly, our results identify Zeeman-induced Fermi surface splitting as a mechanism for the development of nodal excitations in multiband systems. This mechanism should be relevant to a wider class of superconductors and may help interpret field-tuned superconducting phase diagrams in related materials.

\textit{Acknowledgments:} We thank Kazi Ranjibul Islam and Yue Yu for useful discussions. The work of EKK  and AVC was supported by the National Science Foundation grant NSF: DMR-2325357. The work of JJB has been supported by the UK Engineering Physical Sciences Research Council (EPSRC) through grants EP/T034351/1 and EP/X012557/1. EKK acknowledges additional  cost-of-living support by the Onassis Foundation Scholarship ID: F ZU 034-1/2024-2025.

\bibliography{PdBi2}
\appendix
\newpage
\vspace*{-2cm}
{\centering\section{Pairing interaction Hamiltonian in the band basis}
\label{AppendixA}}

In this Appendix we present the explicit form of the pairing interaction Hamiltonian in the band basis. It is convenient to rewrite the interaction Hamiltonian at $\textbf{q}=0$ in the original basis as:
\begin{align}
\mathcal{H}_{int} & =\frac{U+V}{2}\sum_{\textbf{k},\textbf{p}}\left[\sum_{s_{1}}\left(a_{\textbf{k},s_{1}}^{\dagger}a_{\textbf{p},s_{1}}+b_{\textbf{k},s_{1}}^{\dagger}b_{\textbf{p},s_{1}}\right)\right]\left[\sum_{s_{2}}\left(a_{-\textbf{k},s_{2}}^{\dagger}a_{-\textbf{p},s_{2}}+b_{-\textbf{k},s_{2}}^{\dagger}b_{-\textbf{p},s_{2}}\right)\right]\nonumber \\
 & +\frac{U-V}{2}\sum_{\textbf{k},\textbf{p}}\left[\sum_{s_{1}}\left(a_{\textbf{k},s_{1}}^{\dagger}a_{\textbf{p},s_{1}}-b_{\textbf{k},s_{1}}^{\dagger}b_{\textbf{p},s_{1}}\right)\right]\left[\sum_{s_{2}}\left(a_{-\textbf{k},s_{2}}^{\dagger}a_{-\textbf{p},s_{2}}-b_{-\textbf{k},s_{2}}^{\dagger}b_{-\textbf{p},s_{2}}\right)\right] \nonumber\\
 & =\frac{U+V}{2}\sum_{\textbf{k},\textbf{p}}\left[\sum_{s_{1}}\left(c_{\textbf{k},s_{1}}^{\dagger}c_{\textbf{p},s_{1}}+d_{\textbf{k},s_{1}}^{\dagger}d_{\textbf{p},s_{1}}\right)\right]\left[\sum_{s_{2}}\left(c_{-\textbf{k},s_{2}}^{\dagger}c_{-\textbf{p},s_{2}}+d_{-\textbf{k},s_{2}}^{\dagger}d_{-\textbf{p},s_{2}}\right)\right]\nonumber\\
 & +\frac{U-V}{2}\sum_{\textbf{k},\textbf{p}}\left[\sum_{s_{1}}\left(c_{\textbf{k},s_{1}}^{\dagger}d_{\textbf{p},s_{1}}+d_{\textbf{k},s_{1}}^{\dagger}c_{\textbf{p},s_{1}}\right)\right]\left[\sum_{s_{2}}\left(c_{-\textbf{k},s_{2}}^{\dagger}d_{-\textbf{p},s_{2}}+d_{-\textbf{k},s_{2}}^{\dagger}c_{-\textbf{p},s_{2}}\right)\right]
\end{align}
where we recall that
\begin{align}
c_{\textbf{k},s}&=\frac{a_{\textbf{k},s}+b_{\textbf{k},s}}{\sqrt{2}}\nonumber \\
d_{\textbf{k},s}&=\frac{a_{\textbf{k},s}-b_{\textbf{k},s}}{\sqrt{2}}
\end{align}
Using Eq. (\ref{eq:6}) one finds in the band basis that
\begin{align}
\mathcal{H}_{int} & =(U+V)\sum_{\textbf{k},\textbf{p}}\gamma_{k}\gamma_{p}z_{k}z_{p}e^{i\left(\varphi_{p}-\varphi_{k}\right)}e_{\textbf{k}}^{\dagger}e_{-\textbf{k}}^{\dagger}e_{-\textbf{p}}e_{\textbf{p}}-(U+V)\sum_{\textbf{k},\textbf{p}}\gamma_{k}\gamma_{p}z_{k}z_{p}e^{-i\left(\varphi_{k}+\varphi_{p}\right)}e_{\textbf{k}}^{\dagger}e_{-\textbf{k}}^{\dagger}f_{-\textbf{p}}f_{\textbf{p}}\nonumber \\
 & +(U+V)\sum_{\textbf{k},\textbf{p}}\gamma_{k}\gamma_{p}z_{k}z_{p}e^{i\left(\varphi_{k}-\varphi_{p}\right)}f_{\textbf{k}}^{\dagger}f_{-\textbf{k}}^{\dagger}f_{-\textbf{p}}f_{\textbf{p}}+(U+V)\sum_{\textbf{k},\textbf{p}}\gamma_{k}\gamma_{p}z_{k}z_{p}e^{i\left(\varphi_{k}-\varphi_{p}\right)}\bar{e}_{\textbf{k}}^{\dagger}\bar{e}_{-\textbf{k}}^{\dagger}\bar{e}_{-\textbf{p}}\bar{e}_{\textbf{p}}\nonumber \\
 & -(U+V)\sum_{\textbf{k},\textbf{p}}\gamma_{k}\gamma_{p}z_{k}z_{p}e^{i\left(\varphi_{k}+\varphi_{p}\right)}\bar{e}_{\textbf{k}}^{\dagger}\bar{e}_{-\textbf{k}}^{\dagger}\bar{f}_{-\textbf{p}}\bar{f}_{\textbf{p}}+(U+V)\sum_{\textbf{k},\textbf{p}}\gamma_{k}\gamma_{p}z_{k}z_{p}e^{i\left(\varphi_{p}-\varphi_{k}\right)}\bar{f}_{\textbf{k}}^{\dagger}\bar{f}_{-\textbf{k}}^{\dagger}\bar{f}_{-\textbf{p}}\bar{f}_{\textbf{p}}\nonumber \\
 & +(U-V)\sum_{\textbf{k},\textbf{p}}\gamma_{k}\gamma_{p}z_{k}z_{p}e^{-i\left(\varphi_{k}+\varphi_{p}\right)}e_{\textbf{k}}^{\dagger}e_{-\textbf{k}}^{\dagger}\bar{e}_{-\textbf{p}}\bar{e}_{\textbf{p}}-(U-V)\sum_{\textbf{k},\textbf{p}}\gamma_{k}\gamma_{p}z_{k}z_{p}e^{i\left(\varphi_{p}-\varphi_{k}\right)}e_{\textbf{k}}^{\dagger}e_{-\textbf{k}}^{\dagger}\bar{f}_{-\textbf{p}}\bar{f}_{\textbf{p}}\nonumber \\
 & -(U-V)\sum_{\textbf{k},\textbf{p}}\gamma_{k}\gamma_{p}z_{k}z_{p}e^{i\left(\varphi_{k}-\varphi_{p}\right)}f_{\textbf{k}}^{\dagger}f_{-\textbf{k}}^{\dagger}\bar{e}_{-\textbf{p}}\bar{e}_{\textbf{p}}+(U-V)\sum_{\textbf{k},\textbf{p}}\gamma_{k}\gamma_{p}z_{k}z_{p}e^{i\left(\varphi_{k}+\varphi_{p}\right)}f_{\textbf{k}}^{\dagger}f_{-\textbf{k}}^{\dagger}\bar{f}_{-\textbf{p}}\bar{f}_{\textbf{p}}\nonumber \\
 & +\sum_{\textbf{k},\textbf{p}}\Bigl\{(U+V)\left[(\gamma_k \gamma_p)^2+(z_k z_p)^2\right]+(U-V)\left[(\gamma_k z_p)^2+(z_k \gamma_p)^2\right]\Bigr\}e_{\textbf{k}}^{\dagger}\bar{e}_{-\textbf{k}}^{\dagger}\bar{e}_{-\textbf{p}}e_{\textbf{p}} \nonumber\\
 & -\sum_{\textbf{k},\textbf{p}}\Bigl\{(U-V)\left[(\gamma_k \gamma_p)^2+(z_k z_p)^2\right]+(U+V)\left[(\gamma_k z_p)^2+(z_k \gamma_p)^2\right]\Bigr\}e_{\textbf{k}}^{\dagger}\bar{e}_{-\textbf{k}}^{\dagger}\bar{f}_{-\textbf{p}}f_{\textbf{p}}\nonumber \\
 & +\sum_{\textbf{k},\textbf{p}}\Bigl\{(U+V)\left[(\gamma_k \gamma_p)^2+(z_k z_p)^2\right]+(U-V)\left[(\gamma_k z_p)^2+(z_k \gamma_p)^2\right]\Bigr\}f_{\textbf{k}}^{\dagger}\bar{f}_{-\textbf{k}}^{\dagger}\bar{f}_{-\textbf{p}}f_{\textbf{p}}+\text{h.c.}
 \label{eq:1.15}
\end{align}
where we kept only the terms of the interaction that can contribute
to superconductivity. Since we are dealing with effectively pseudospin operators, we only kept the triplet component of the interaction for the terms that promote pairing between fermions of the same type.

\begin{figure}[H]
\begin{center}
\includegraphics[width=1\textwidth]{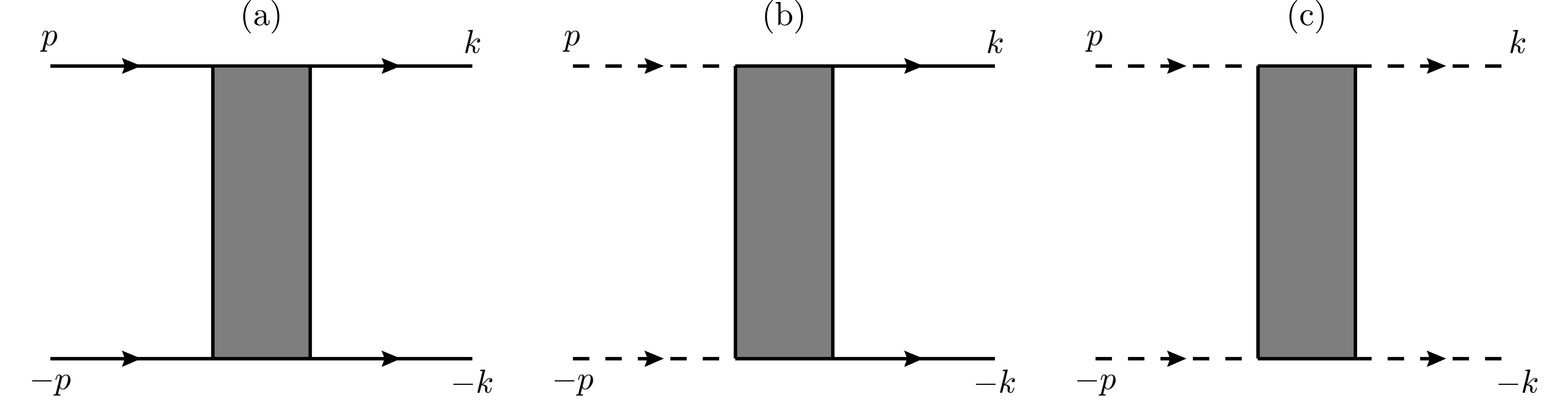}
\end{center}
\caption{Diagrammatic representation of the irreducible pairing vertices corresponding to (a) $e_k^\dagger e_{-k}^\dagger e_{-p}e_p$, (b) $e_k^\dagger e_{-k}^\dagger f_{-p}f_p$ and (c) $f_k^\dagger f_{-k}^\dagger f_{-p}f_p$. The solid line represents $e$ fermions, while the dashed line represents $f$ fermions.}
\label{fig:10}
\end{figure}

\vspace{5mm}
{\centering\section{Kohn-Luttinger corrections to the interaction}
\label{AppendixB}}
In this Appendix, we treat the interaction components in Eqs. (\ref{eq:17}) and (\ref{eq:22}), which are equal in magnitude at the bare level, as irreducible in the pairing channel and dress them by particle-hole corrections. The relevant pairing interaction components for the p-wave channel are shown in Fig. \ref{fig:10}. 
Since we are including Kohn-Luttinger type renormalizations from the paricle-hole channel, we can no longer project the interaction Hamiltonian onto the pairing channel alone and must instead use the full interaction Hamiltonian
\begin{equation}
\mathcal{H}_{int}=U\sum_{\substack{\mathbf{k},\mathbf{p},\mathbf{q} \\ s_{1},s_{2}}}\left[a_{\textbf{k},s_{1}}^{\dagger}a_{\textbf{p},s_{1}}a_{-\textbf{k}+\textbf{q},s_{2}}^{\dagger}a_{-\textbf{p}+\textbf{q},s_{2}}+b_{\textbf{k},s_{1}}^{\dagger}b_{\textbf{p},s_{1}}b_{-\textbf{k}+\textbf{q},s_{2}}^{\dagger}b_{-\textbf{p}+\textbf{q},s_{2}}\right]
\end{equation}
where for simplicity we set $V=0$ for this calculation. In the $e$-$f$ fermion basis, this interaction Hamiltonian can be written as
\begin{equation}
\mathcal{H}_{int}=\frac{U}{2}\sum_{\mathbf{k},\mathbf{p},\mathbf{q}}\left[N_{1,\textbf{k},\textbf{p}}N_{1,-\textbf{k}+\textbf{q},-\textbf{p}+\textbf{q}}+N_{2,\textbf{k},\textbf{p}}N_{2,-\textbf{k}+\textbf{q},-\textbf{p}+\textbf{q}}\right]
\end{equation}
\begin{figure}[H]
\begin{center}
\includegraphics[width=1\textwidth]{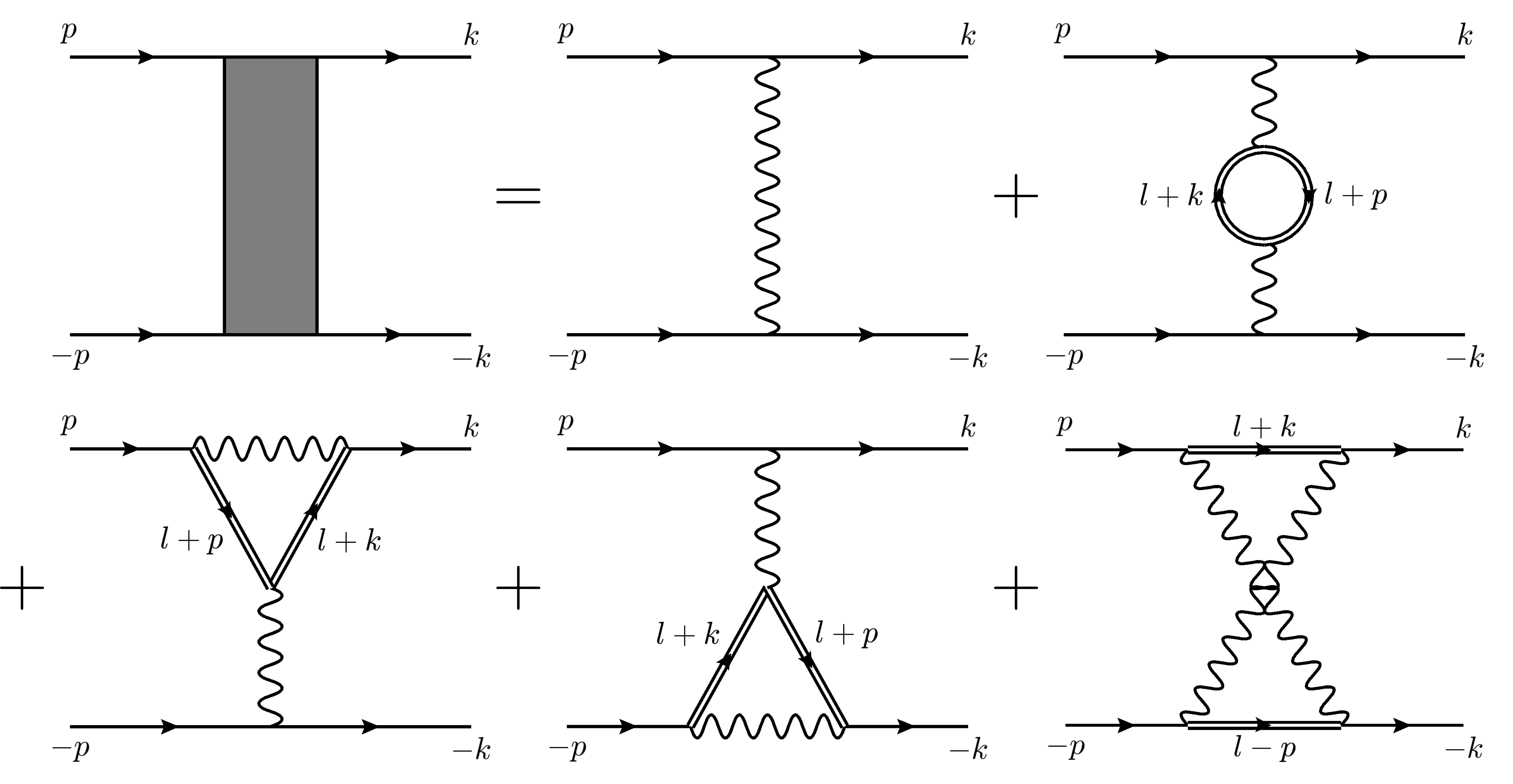}
\end{center}
\caption{The irreducible pairing vertex $e_k^\dagger e_{-k}^\dagger e_{-p}e_p$ (the diagrams for the other components are directly analogous) to second order in the interaction $U$, depicted by the wavy lines. In this notation, the double line propagators represent an internal summation over all possible fermionic flavors, such that the total number of barred and unbarred fermions in each four-fermion vertex is even.}
\label{fig:11}
\end{figure}
where we have defined the density operators
\begin{align}
N_{1,\textbf{k},\textbf{p}}=&+\Bigl[\gamma_k\gamma_p+e^{i(\varphi_p-\varphi_k)}z_kz_p\Bigr]e^\dagger_k e_p+\Bigl[i\gamma_k z_pe^{-i\varphi_p}-i\gamma_p z_k e^{-i\varphi_k}\Bigr]e^\dagger_k f_p\nonumber \\
&+\Bigl[i\gamma_kz_pe^{i\varphi_p}-i\gamma_pz_ke^{i\varphi_k}\Bigr]f^\dagger_ke_p+\Bigl[\gamma_k\gamma_p+e^{i(\varphi_k-\varphi_p)}z_kz_p\Bigr]f^\dagger_kf_p\nonumber \\
&+\Bigl[\gamma_k\gamma_p+e^{i(\varphi_k-\varphi_p)}z_kz_p\Bigr]\bar{e}^\dagger_k\bar{e}_p+\Bigl[i\gamma_pz_ke^{i\varphi_k}-i\gamma_kz_pe^{i\varphi_p}\Bigr]\bar{e}^\dagger_k\bar{f}_p\nonumber \\
&+\Bigl[i\gamma_pz_ke^{-i\varphi_k}-i\gamma_kz_pe^{-i\varphi_p}\Bigr]\bar{f}^\dagger_k\bar{e}_p+\Bigl[\gamma_k\gamma_p+e^{i(\varphi_p-\varphi_k)}z_kz_p\Bigr]\bar{f}^\dagger_k\bar{f}_p
\end{align}
\begin{align}
N_{2,\textbf{k},\textbf{p}}=&-\Bigl[i\gamma_kz_pe^{-i\varphi_p}+i\gamma_pz_ke^{-i\varphi_k}\Bigr]e^\dagger_k\bar{e}_p+\Bigl[\gamma_k\gamma_p-e^{i(\varphi_p-\varphi_k)}z_kz_p\Bigr]e^\dagger_k\bar{f}_p\nonumber \\
&+\Bigl[\gamma_k\gamma_p-e^{i(\varphi_k-\varphi_p)}z_kz_p\Bigr]f^\dagger_k \bar{e}_p-\Bigl[i\gamma_k z_pe^{i\varphi_p}+i\gamma_p z_k e^{i\varphi_k}\Bigr]f^\dagger_k \bar{f}_p\nonumber \\
&+\Bigl[i\gamma_kz_pe^{i\varphi_p}+i\gamma_pz_ke^{i\varphi_k}\Bigr]\bar{e}^\dagger_ke_p+\Bigl[\gamma_k\gamma_p-e^{i(\varphi_k-\varphi_p)}z_kz_p\Bigr]\bar{e}^\dagger_kf_p\nonumber \\
&+\Bigl[\gamma_k\gamma_p-e^{i(\varphi_p-\varphi_k)}z_kz_p\Bigr]\bar{f}^\dagger_ke_p+\Bigl[i\gamma_kz_pe^{-i\varphi_p}+i\gamma_pz_ke^{-i\varphi_k}\Bigr]\bar{f}^\dagger_kf_p
\end{align}

\begin{figure}[H]
\begin{center}
\includegraphics[width=1\textwidth]{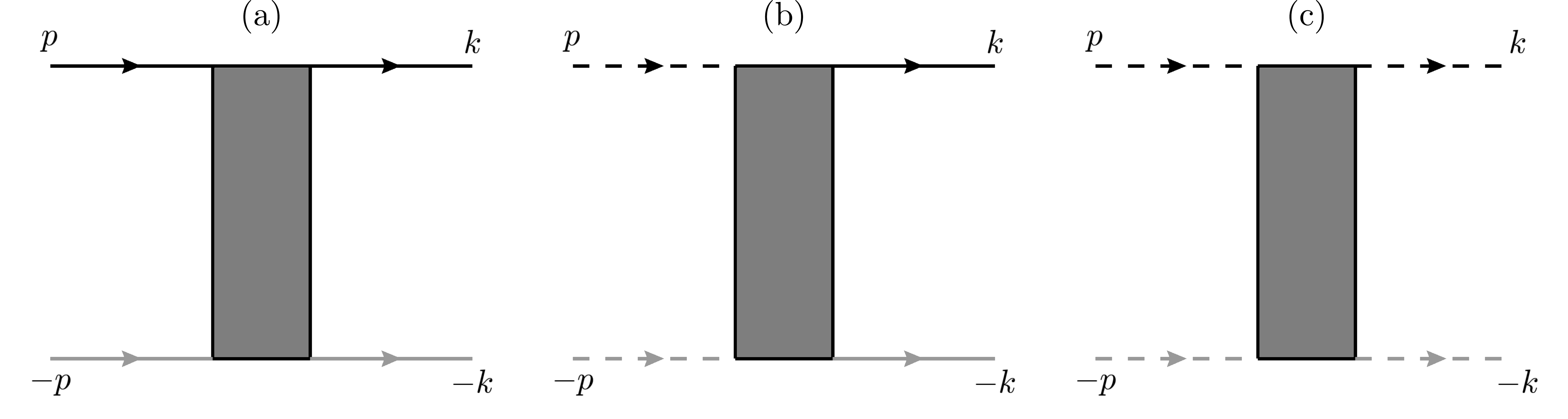}
\end{center}
\caption{Diagrammatic representation of the irreducible pairing vertices corresponding to (a) $e_k^\dagger \bar{e}_{-k}^\dagger \bar{e}_{-p}e_p$, (b) $e_k^\dagger \bar{e}_{-k}^\dagger \bar{f}_{-p}f_p$ and (c) $f_k^\dagger \bar{f}_{-k}^\dagger \bar{f}_{-p}f_p$. The solid lines represent $e$ fermions, while the dashed lines represent $f$ fermions, with the black lines corresponding to unbarred and the grey lines corresponding to barred fermions.}
\label{fig:12}
\end{figure}

One can immediately notice that, in the interaction Hamiltonian, each term contains an even number of barred and unbarred fermions. We can now proceed with the calculation of higher-order corrections to the interaction. Those are shown in Fig. \ref{fig:11}, where in our notation, the double line propagators imply a summation over all fermionic flavors such that the previous condition is met. We note here that antisymmetrized diagrams do not need to be calculated separately as they simply impose that we keep only the triplet component of the interaction. Here, we numerically evaluate the contribution of all second-order corrections from the particle hole channel to the irreducible vertices shown in Fig. \ref{fig:10}. The numerical calculation is fairly straightforward but quite cumbersome due to the form factors of the interaction. We find that
\begin{align}
\Gamma^{(2)}_{eeee}(\textbf{k},\textbf{p})&=+\;U\gamma_k\gamma_pz_kz_pe^{i(\varphi_p-\varphi_k)}\Bigl[1+1.957N_FU\Bigr]\nonumber\\
\Gamma^{(2)}_{eeff}(\textbf{k},\textbf{p})&=-\;U\gamma_k\gamma_pz_kz_pe^{-i(\varphi_k+\varphi_p)}\Bigl[1+2.206N_F U\Bigr]\nonumber\\
\Gamma^{(2)}_{ffff}(\textbf{k},\textbf{p})&=+\;U\gamma_k\gamma_pz_kz_pe^{i(\varphi_k-\varphi_p)}\Bigl[1+2.013N_F U\Bigr]
\end{align}
where $N_F=\frac{N_{F+}+N_{F-}}{2}$ and $\textbf{k}$ and $\textbf{p}$ lie on top of the respective Fermi surfaces. We also note here that in our numerics, we project $\Gamma^{(2)}(\textbf{k},\textbf{p})$ into the p-wave channel and ignore all further corrections beyond it. The renormalized components $U_1$ and $U_2$ are given by
\begin{align}
U_1&=U\Bigl[1+1.985N_FU\Bigr]\nonumber\\
U_2&=U\Bigl[1+2.206N_FU\Bigr]
\end{align}
where $U_1$ was approximated by taking the average of the two intrapocket components of the interaction. As previously advertised, we find that the interpocket vertex is enhanced compared to the intrapocket one, similar to the case of the cuprates and iron-based superconductors. We now move on to the s-wave channel. The relevant pairing interaction components for the s-wave channel are shown in Fig. \ref{fig:12}. The second order diagrams are the same as before, with the difference that for the s-wave channel, antisymmetrized diagrams need to be explicitly evaluated. Numerically evaluating the second order corrections in this channel, we find
\begin{align}
\Gamma^{(2)}_{e\bar{e}\bar{e}e}(\textbf{k},\textbf{p})&=+\;U\Bigl[1+1.979N_FU\Bigr]\nonumber\\
\Gamma^{(2)}_{e\bar{e}\bar{f}f}(\textbf{k},\textbf{p})&=-\;U\Bigl[1+1.963N_F U\Bigr]\nonumber\\
\Gamma^{(2)}_{f\bar{f}\bar{f}f}(\textbf{k},\textbf{p})&=+\;U\Bigl[1+1.995N_F U\Bigr]
\label{eq:B7}
\end{align}
One can immediately notice that the second order Kohn-Luttinger corrections in the s-wave channel do not by themselves generate a net attractive pairing interaction. As shown in Eq. (\ref{eq:B7}), the renormalized interpocket interaction is slightly smaller than the corresponding intrapocket interactions at this order. Thus, unlike in the p-wave channel, the second-order calculation does not lift the cancellation between the bare attractive interpocket and repulsive intrapocket contributions in favor of superconductivity.

\newpage
{\centering\section{Linearized gap equation in the s-wave channel at h=0}
\label{AppendixC}}
In this Appendix, we present an analytical treatment of the linearized gap equation for the s-wave interband pairing channel at $h=0$. The linearized gap equation for the pairings $\Delta_{e\bar{e}}\left(\textbf{k}\right)$ and $\Delta_{f\bar{f}}\left(\textbf{k}\right)$ is given by
\begin{equation}
\Delta_{\alpha\bar{\alpha}}\left(\textbf{k}\right)=-\sum_{\beta} \intop\frac{d^{2}\textbf{p}}{\left(2\pi\right)^{2}}\frac{U^{\text{inter}}_{\alpha\bar{\alpha};\bar{\beta}\beta}(\textbf{k},\textbf{p})\Delta_{\beta\bar{\beta}}\left(\textbf{p}\right)}{2\varepsilon_{\beta}(\textbf{p})}\tanh\left(\frac{\varepsilon_{\beta}(\textbf{p})}{2T_c}\right)
\end{equation}
where $\alpha$ and $\beta$ are indices that run over $e$ and $f$ only. This system of integral equations can be simplified by assuming that $\Delta_{e\bar{e}}(\textbf{k})=\Delta_1 \gamma_k^2+\Delta_2z_k^2$ and $\Delta_{f\bar{f}}(\textbf{k})=\Delta_3 \gamma_k^2+\Delta_4z_k^2$. Substituting this Ansatz into the linearized gap equation we obtain: 
\begin{align}
\begin{pmatrix}
\Delta_1 \\
\Delta_2 \\
\Delta_3 \\
\Delta_4
\end{pmatrix}
&=\frac{U+V}{2}
\begin{pmatrix}
-L_{\gamma\gamma,+} & -L_{\gamma z,+} & L_{\gamma z,-} & L_{zz,-} \\
-L_{\gamma z,+} & -L_{zz,+} & L_{\gamma\gamma,-} & L_{\gamma z,-} \\
L_{\gamma z,+} & L_{zz,+} & -L_{\gamma\gamma,-} & -L_{\gamma z,-} \\
L_{\gamma\gamma,+} & L_{\gamma z,+} & -L_{\gamma z,-} & -L_{zz,-}
\end{pmatrix}
\begin{pmatrix}
\Delta_1 \\
\Delta_2 \\
\Delta_3 \\
\Delta_4
\end{pmatrix}\nonumber\\
&+\frac{U-V}{2}
\begin{pmatrix}
-L_{\gamma z,+} & -L_{zz,+} & L_{\gamma\gamma,-} & L_{\gamma z,-} \\
-L_{\gamma\gamma,+} & -L_{\gamma z,+} & L_{\gamma z,-} & L_{zz,-} \\
L_{\gamma\gamma,+} & L_{\gamma z,+} & -L_{\gamma z,-} & -L_{zz,-} \\
L_{\gamma z,+} & L_{zz,+} & -L_{\gamma\gamma,-} & -L_{\gamma z,-} 
\end{pmatrix}
\begin{pmatrix}
\Delta_1 \\
\Delta_2 \\
\Delta_3 \\
\Delta_4
\end{pmatrix}
\label{eq:B2}
\end{align}
where:
\begin{align}
L_{\gamma\gamma,\pm}&=\intop\frac{d^{2}\textbf{p}}{\left(2\pi\right)^{2}}\frac{\gamma_{p}^{4}}{\varepsilon_{\pm}\left(p\right)}\tanh\left(\frac{\varepsilon_{\pm}\left(p\right)}{2T_{c}}\right)\approx 2N_{F\pm}\gamma_{k_{F\pm}}^4\log\frac{1.13\omega_D}{T_c} \nonumber\\
L_{\gamma z,\pm}&=\intop\frac{d^{2}\textbf{p}}{\left(2\pi\right)^{2}}\frac{(\gamma_{p}z_p)^{2}}{\varepsilon_{\pm}\left(p\right)}\tanh\left(\frac{\varepsilon_{\pm}\left(p\right)}{2T_{c}}\right)\approx 2N_{F\pm}\gamma_{k_{F\pm}}^2z_{k_{F\pm}}^2\log\frac{1.13\omega_D}{T_c}\nonumber\\
L_{zz,\pm}&=\intop\frac{d^{2}\textbf{p}}{\left(2\pi\right)^{2}}\frac{z_{p}^{4}}{\varepsilon_{\pm}\left(p\right)}\tanh\left(\frac{\varepsilon_{\pm}\left(p\right)}{2T_{c}}\right)\approx 2N_{F\pm}z_{k_{F\pm}}^4\log\frac{1.13\omega_D}{T_c}
\end{align}
This equation again does not have a solution unless the interactions are renormalized as before. For simplicity, let us adopt a renormalization of the form, $U_{1/2}=U(1\mp\epsilon)$ and similarly $V_{1/2}=V(1\mp\epsilon)$, characterized by a single parameter $\epsilon$. In this notation, $U_1$ and $V_1$ correspond to the block diagonal repulsive components of the gap equation, while $U_2$ and $V_2$ correspond to the block off-diagonal attractive components. Looking for a non trivial solution of Eq. (\ref{eq:B2}), we find after some algebra that the parameter $\epsilon$ satisfies the equation
\begin{equation}
a-b\epsilon+c\epsilon^2=0
\label{eq:B4}
\end{equation}
where for the parameters presented in Sec. \ref{secII}
\begin{align}
a&\approx1+\left(0.138\cdot U+0.036\cdot V\right)\log{\frac{1.13\omega_D}{T_c}}+0.005\cdot UV\log^2{\frac{1.13\omega_D}{T_c}}      \nonumber \\
b&\approx\left(0.138\cdot U+0.036\cdot V\right)\log{\frac{1.13\omega_D}{T_c}}+\left(0.018\cdot U^2+0.0007\cdot UV+0.0013\cdot V^2\right)\log^2{\frac{1.13\omega_D}{T_c}}\nonumber \\
c&\approx0.005\cdot UV\log^2{\frac{1.13\omega_D}{T_c}}
\label{eq:B5}
\end{align}
One can immediately notice that $V$ does not drop out of the gap equation in this channel as it is present in Eq. (\ref{eq:B5}). We also note that for nonzero $U$ and $V$, the prefactor of $\epsilon^2$ does not vanish. Taking the limit $T_c\rightarrow0$, we find that $\epsilon$ satisfies the equation
\begin{equation}
\epsilon^2-\left[3.6\cdot\frac{U}{V}+0.14+0.26\cdot\frac{V}{U}\right]\epsilon+1=0
\end{equation}
This equation has a real finite solution $\epsilon_c<1$ for all values of $U/V$. In other words, superconductivity in this channel develops only for $\epsilon>\epsilon_c$. The reason is that all three coefficients in Eq. (\ref{eq:B4}) diverge as $\log^2{T_c}$ in the limit $T_c\rightarrow0$. However, when either $U=0$ or $V=0$, only $b$ continues to exhibit a $\log^2{T_c}$ divergence, while $c$ vanishes and $a$ diverges only logarithmically, $a\sim\log{T_c}$. In this case, $\epsilon_c=0$ and superconductivity develops for all values of $\epsilon$ as was the case in the p-wave channel. In the case where $V=0$, one finds by solving Eq. (\ref{eq:B4})
\begin{equation}
\epsilon_{s\text{-wave}}\approx\frac{7.7}{U\log{\frac{1.13\omega_D}{T_c}}}
\end{equation}
As expected, $\epsilon_{s\text{-wave}}\rightarrow0$ when $T_c\rightarrow0$. Applying the same interaction renormalization scheme in the case of the p-wave channel (Eq. \ref{eq:Tc1}), we find
\begin{equation}
\epsilon_{p\text{-wave}}\approx\frac{11.4}{U\log{\frac{1.13\omega_D}{T_c}}}
\end{equation}
Assuming that the renormalization factors $\epsilon$ are the same, or at least close, for both channels, we see that $T_{c,s\text{-wave}}>T_{c,p\text{-wave}}$ in agreement with our numerical results in Sec. \ref{secIIIB}.

\newpage
{\centering\section{Approximate analysis of the gap equations in the p-wave channel}
\label{AppendixD}}

In this Appendix, we adopt an approximate form for the unitary transformation matrix $A(\textbf{k})$ and investigate the solutions of the gap equations for the p-wave channel for different values of the magnetic field. Specifically, we assume
\begin{equation}
\psi_{0,\textbf{k}}^\dagger=\tfrac{1}{\sqrt{2}}\begin{pmatrix}
ie^{i\phi_k} & -ie^{i\phi_k} & 0 & 0 \\
-ie^{-i\phi_k} & -ie^{-i\phi_k} & 0 & 0 \\
0 & 0 & ie^{-i\phi_k} & ie^{-i\phi_k} \\
0 & 0 & ie^{i\phi_k} & -ie^{i\phi_k}
\end{pmatrix}\psi_\textbf{k}^\dagger
\end{equation}
Using Eqs. (\ref{eq:38}) and (\ref{eq:39}), we can then derive the linearized gap equations. At zero field, the linearized gap equations take the form
\begin{align}
\Delta_{gg}\left(\textbf{k}\right)&=\intop\frac{d^{2}\textbf{p}}{\left(2\pi\right)^{2}}\gamma_{k}\gamma_{p}z_{k}z_{p}\Bigl[(U+V)\cos{(\varphi_k+\varphi_p)+(U-V)\cos{(\varphi_k-\varphi_p)}}\Bigl]\Delta_{jj}\left(\textbf{p}\right)\frac{\tanh\left(\frac{\varepsilon_-(p)}{2T_c}\right)}{\varepsilon_{-}\left(p\right)}\nonumber \\
 & +\intop\frac{d^{2}\textbf{p}}{\left(2\pi\right)^{2}}\gamma_{k}\gamma_{p}z_{k}z_{p}\Bigl[i(U+V)\sin{(\varphi_k+\varphi_p)-i(U-V)\sin{(\varphi_k-\varphi_p)}}\Bigl]\Delta_{j\bar{j}}\left(\textbf{p}\right)\frac{\tanh\left(\frac{\varepsilon_-(p)}{2T_c}\right)}{\varepsilon_{-}\left(p\right)}\nonumber \\
\Delta_{g\bar{g}}\left(\textbf{k}\right)&=\intop\frac{d^{2}\textbf{p}}{\left(2\pi\right)^{2}}\gamma_{k}\gamma_{p}z_{k}z_{p}\Bigl[-i(U+V)\sin{(\varphi_k+\varphi_p)-i(U-V)\sin{(\varphi_k-\varphi_p)}}\Bigl]\Delta_{jj}\left(\textbf{p}\right)\frac{\tanh\left(\frac{\varepsilon_-(p)}{2T_c}\right)}{\varepsilon_{-}\left(p\right)}\nonumber \\
 & +\intop\frac{d^{2}\textbf{p}}{\left(2\pi\right)^{2}}\gamma_{k}\gamma_{p}z_{k}z_{p}\Bigl[-(U+V)\cos{(\varphi_k+\varphi_p)+(U-V)\cos{(\varphi_k-\varphi_p)}}\Bigl]\Delta_{j\bar{j}}\left(\textbf{p}\right)\frac{\tanh\left(\frac{\varepsilon_-(p)}{2T_c}\right)}{\varepsilon_{-}\left(p\right)}\nonumber \\
\Delta_{jj}\left(\textbf{k}\right)&=\intop\frac{d^{2}\textbf{p}}{\left(2\pi\right)^{2}}\gamma_{k}\gamma_{p}z_{k}z_{p}\Bigl[(U+V)\cos{(\varphi_k+\varphi_p)+(U-V)\cos{(\varphi_k-\varphi_p)}}\Bigl]\Delta_{gg}\left(\textbf{p}\right)\frac{\tanh\left(\frac{\varepsilon_+(p)}{2T_c}\right)}{\varepsilon_{+}\left(p\right)}\nonumber \\
 & +\intop\frac{d^{2}\textbf{p}}{\left(2\pi\right)^{2}}\gamma_{k}\gamma_{p}z_{k}z_{p}\Bigl[i(U+V)\sin{(\varphi_k+\varphi_p)-i(U-V)\sin{(\varphi_k-\varphi_p)}}\Bigl]\Delta_{g\bar{g}}\left(\textbf{p}\right)\frac{\tanh\left(\frac{\varepsilon_+(p)}{2T_c}\right)}{\varepsilon_{+}\left(p\right)}\nonumber \\
\Delta_{j\bar{j}}\left(\textbf{k}\right)&=\intop\frac{d^{2}\textbf{p}}{\left(2\pi\right)^{2}}\gamma_{k}\gamma_{p}z_{k}z_{p}\Bigl[-i(U+V)\sin{(\varphi_k+\varphi_p)-i(U-V)\sin{(\varphi_k-\varphi_p)}}\Bigl]\Delta_{gg}\left(\textbf{p}\right)\frac{\tanh\left(\frac{\varepsilon_+(p)}{2T_c}\right)}{\varepsilon_{+}\left(p\right)}\nonumber \\
 & +\intop\frac{d^{2}\textbf{p}}{\left(2\pi\right)^{2}}\gamma_{k}\gamma_{p}z_{k}z_{p}\Bigl[-(U+V)\cos{(\varphi_k+\varphi_p)+(U-V)\cos{(\varphi_k-\varphi_p)}}\Bigl]\Delta_{g\bar{g}}\left(\textbf{p}\right)\frac{\tanh\left(\frac{\varepsilon_+(p)}{2T_c}\right)}{\varepsilon_{+}\left(p\right)}
\end{align}
where for simplicity, we have completely suppressed the intrapocket (block-diagonal) terms. We note that, due to the structure of the interaction in the $g$ and $j$ fermion basis, the gap functions satisfy the relations $\Delta_{gg}(\textbf{k})=\Delta_{\bar{g}\bar{g}}(\textbf{k})$ and $\Delta_{jj}(\textbf{k})=\Delta_{\bar{j}\bar{j}}(\textbf{k})$. The above equation admits a solution of the form $\Delta_{gg}(\textbf{k})=\Delta_{gg}  \gamma_kz_k\cos{\varphi_k}$, $\Delta_{g\bar{g}}(\textbf{k})=-i\Delta_{g\bar{g}}\gamma_kz_k\sin{\varphi_k}$, $\Delta_{jj}(\textbf{k})=\Delta_{jj} \gamma_kz_k\cos{\varphi_k}$ and $\Delta_{j\bar{j}}(\textbf{k})=-i\Delta_{j\bar{j}}\gamma_kz_k\sin{\varphi_k}$. This turns the integral equation into the following algebraic one
\begin{equation}
\begin{pmatrix}
0 & 0 & \frac{UL_-}{2} & \frac{UL_-}{2} \\
0 & 0 & \frac{UL_-}{2} & \frac{UL_-}{2} \\
\frac{UL_+}{2} & \frac{UL_+}{2} & 0 & 0 \\
\frac{UL_+}{2} & \frac{UL_+}{2} & 0 & 0
\end{pmatrix}
\begin{pmatrix}
\Delta_{gg} \\ \Delta_{g\bar{g}} \\ \Delta_{jj} \\ \Delta_{j\bar{j}}
\end{pmatrix}=
\begin{pmatrix}
\Delta_{gg} \\ \Delta_{g\bar{g}} \\ \Delta_{jj} \\ \Delta_{j\bar{j}}
\end{pmatrix}
\end{equation}
where $L_+$ and $L_-$ are defined in the same way as in Eq. (\ref{eq:3.6}). This system of algebraic equations admits a nonzero solution when the condition $U^2L_+L_-=1$ is met, matching our result for $T_c$ from Eq. (\ref{eq:17}) with the absence of the diagonal elements. A quick analysis of the eigenvectors of this solution gives $\Delta_{gg}=\Delta_{g\bar{g}}$ and $\Delta_{jj}=\Delta_{j\bar{j}}$. However, this solution is not unique. A second solution of the form $\Delta_{gg}(\textbf{k})=-\Delta_{gg}  \gamma_kz_k\sin{\varphi_k}$, $\Delta_{g\bar{g}}(\textbf{k})=-i\Delta_{g\bar{g}}\gamma_kz_k\cos{\varphi_k}$, $\Delta_{jj}(\textbf{k})=\Delta_{jj} \gamma_kz_k\sin{\varphi_k}$ and $\Delta_{j\bar{j}}(\textbf{k})=i\Delta_{j\bar{j}}\gamma_kz_k\cos{\varphi_k}$ exists, reducing the system of integral equation to the same system of algebraic ones as before but now with $V$ instead of $U$. This is fully consistent with our previous analysis in section \ref{secIIIB}. The first solution corresponds to the case where $\Delta_{ee}(\textbf{k})$ and $\Delta_{\bar{e}\bar{e}}(\textbf{k})$ develop with the same sign and has a higher $T_c$ when $U>V$ and the second solution corresponds to the case where the develop with the opposite one and has a higher $T_c$ when $U<V$. However, the case $U=V$ becomes interesting. Since both solutions develop at the same $T_c$ all linear combinations of them solve the linearized gap equation. However, as we will show not all of them survive when considering the non linear gap equation. At non-linear order, the zero magnetic field gap equations in the $g,j$ basis with $U=V$ are given by
\begin{align}
\Delta_{gg}\left(\textbf{k}\right)-\Delta_{g\bar{g}}(\textbf{k})&=2U\intop\frac{d^{2}\textbf{p}}{\left(2\pi\right)^{2}}\gamma_{k}\gamma_{p}z_{k}z_{p}e^{i(\varphi_k+\varphi_p)}\Bigl[\Delta_{jj}(\textbf{p})+\Delta_{j\bar{j}}(\textbf{p})\Bigl]\frac{\tanh\left(\frac{E_{j+}(\textbf{p})}{2T}\right)}{E_{j+}(\textbf{p})}\nonumber \\
\Delta_{gg}\left(\textbf{k}\right)+\Delta_{g\bar{g}}(\textbf{k})&=2U\intop\frac{d^{2}\textbf{p}}{\left(2\pi\right)^{2}}\gamma_{k}\gamma_{p}z_{k}z_{p}e^{-i(\varphi_k+\varphi_p)}\Bigl[\Delta_{jj}(\textbf{p})-\Delta_{j\bar{j}}(\textbf{p})\Bigl]\frac{\tanh\left(\frac{E_{j-}(\textbf{p})}{2T}\right)}{E_{j-}(\textbf{p})}
\end{align}
where we have defined 
\begin{equation}
E_{j\pm}(\textbf{k})=\sqrt{\left(\varepsilon_-(k)\right)^2+|\Delta_{jj}(\textbf{k})|^2+|\Delta_{j\bar{j}}(\textbf{k})|^2\pm\left(\Delta^*_{jj}(\textbf{k})\Delta_{j\bar{j}}(\textbf{k})+\Delta_{jj}(\textbf{k})\Delta^*_{j\bar{j}}(\textbf{k})\right)}
\end{equation}
The equations for $\Delta_{jj}(\textbf{k})$ and $\Delta_{j\bar{j}}(\textbf{k})$ follow in a similar way but we won't need them for our present analysis. Let us search for a solution for the system of integral equations based on the general form of the solutions of the linearized gap equations.
\begin{align}
\Delta_{gg}(\textbf{k})&=\Delta_{gg}\gamma_kz_k(c_1\cos\varphi_k-c_2\sin\varphi_k)\nonumber\\
\Delta_{g\bar{g}}(\textbf{k})&=-i\Delta_{gg}\gamma_kz_k(c_1\sin\varphi_k+c_2\cos\varphi_k)\nonumber\\
\Delta_{jj}(\textbf{k})&=\Delta_{jj}\gamma_kz_k(c_1\cos\varphi_k+c_2\sin\varphi_k)\nonumber\\
\Delta_{j\bar{j}}(\textbf{k})&=-i\Delta_{jj}\gamma_kz_k(c_1\sin\varphi_k-c_2\cos\varphi_k)
\end{align}
Substituting this Ansatz into the nonlinear gap equations we find 
\begin{align}
\Delta_{gg}(c_1+ic_2)&=2U(c_1+ic_2)\Delta_{jj}\intop\frac{d^{2}\textbf{p}}{\left(2\pi\right)^{2}}(\gamma_pz_p)^2\frac{\tanh\left(\frac{E_{j+}(\textbf{p})}{2T}\right)}{E_{j+}(\textbf{p})}\nonumber \\
\Delta_{gg}(c_1-ic_2)&=2U(c_1-ic_2)\Delta_{jj}\intop\frac{d^{2}\textbf{p}}{\left(2\pi\right)^{2}}(\gamma_pz_p)^2\frac{\tanh\left(\frac{E_{j-}(\textbf{p})}{2T}\right)}{E_{j-}(\textbf{p})}
\end{align}

We find two distinct cases. Either $c_1\pm ic_2=0$ has to hold or $E_{j+}(\textbf{k})=E_{j-}(\textbf{k})$ has to hold. The first one corresponds to complex exponential solutions and in reality is an unstable solution of the gap equation since it corresponds to the case where one of $\Delta_{ee}(\textbf{k})$ or $\Delta_{\bar{e}\bar{e}}(\textbf{k})$ spontaneously develops but the other doesn't. The set of true equilibrium solutions are the ones that satisfy $E_{j+}(\textbf{k})=E_{j-}(\textbf{k})$. Those are given by the condition $\text{Im}[c_1^*c_2]=0$ meaning that the coefficients $c_1$ and $c_2$ have to be chosen with the same phase. In turn this leads us to solutions of the form $\Delta_{gg}(\textbf{k})\sim\Delta_{gg}\gamma_kz_k\cos{(\varphi_k+\psi/2)}$ where $\Delta_{gg}(\textbf{k})$ is defined up to some global phase and $\psi$ corresponds to the arbitrary phase difference with which $\Delta_{ee}(\textbf{k})$ and $\Delta_{\bar{e}\bar{e}}(\textbf{k})$ can develop. The spectrum of the BdG Hamiltonian in this basis is now given by
\begin{align}
E_{1}(\textbf{k})&=\sqrt{\left(\varepsilon_{+}(k)\right)^2+|\Delta_{gg}(\textbf{k})|^2+|\Delta_{g{\bar{g}}}(\textbf{k})|^2}\nonumber\\
E_{2}(\textbf{k})&=\sqrt{\left(\varepsilon_{-}(k)\right)^2+|\Delta_{jj}(\textbf{k})|^2+|\Delta_{j\bar{j}}(\textbf{k})|^2}
\end{align}
meaning that the gap is constant at zero magnetic field as one would expect since the gap should stay invariant under a basis transformation. We will now study the gap equations at a finite magnetic field and show that eventually the system develops gapless excitations. Let us start our analysis of the finite magnetic field case with the linearized gap equation. Let us also assume the case $U>V$ to avoid degeneracies. Substituting our previous Ansatz for the solution into the linearized gap equation one finds
\begin{equation}
\begin{pmatrix}
0 & 0 & \frac{UL_{j1}}{2} & \frac{UL_{j2}}{2} \\
0 & 0 & \frac{UL_{j1}}{2} & \frac{UL_{j2}}{2} \\
\frac{UL_{g1}}{2} & \frac{UL_{g2}}{2} & 0 & 0 \\
\frac{UL_{g1}}{2} & \frac{UL_{g2}}{2} & 0 & 0
\end{pmatrix}
\begin{pmatrix}
\Delta_{gg} \\ \Delta_{g\bar{g}} \\ \Delta_{jj} \\ \Delta_{j\bar{j}}
\end{pmatrix}=
\begin{pmatrix}
\Delta_{gg} \\ \Delta_{g\bar{g}} \\ \Delta_{jj} \\ \Delta_{j\bar{j}}
\end{pmatrix}
\end{equation}
where we have defined 
\begin{equation}
L_{g/j1}=2\intop\frac{d^{2}\textbf{p}}{\left(2\pi\right)^{2}}\cos^2{\varphi_p}\left(\gamma_{p}z_{p}\right)^{2}\left(\frac{\tanh\left(\frac{\varepsilon_{g/j}\left(\textbf{p}\right)}{2T_{c}}\right)}{\varepsilon_{g/j}\left(\textbf{p}\right)}+\frac{\tanh\left(\frac{\varepsilon_{\bar{g}/\bar{j}}\left(\textbf{p}\right)}{2T_{c}}\right)}{\varepsilon_{\bar{g}/\bar{j}}\left(\textbf{p}\right)}\right)
\end{equation}
\begin{equation}
L_{g/j2}=4\intop\frac{d^{2}\textbf{p}}{\left(2\pi\right)^{2}}\sin^2{\varphi_p}\left(\gamma_{p}z_{p}\right)^{2}\frac{\tanh\left(\frac{\varepsilon_{g/j}\left(\textbf{p}\right)}{2T_{c}}\right)+\tanh\left(\frac{\varepsilon_{\bar{g}/\bar{j}}\left(\textbf{p}\right)}{2T_{c}}\right)}{\varepsilon_{g/j}\left(\textbf{p}\right)+\varepsilon_{\bar{g}/\bar{j}}\left(\textbf{p}\right)}
\end{equation}
Even though at high enough magnetic fields $L_{g/j1}\gg L_{g/j2}$ because the Cooper logarithm is lost in the case where pairing occurs between fermions in different Fermi surfaces, the solution still develops with $\Delta_{gg}=\Delta_{g\bar{g}}$. For now let us assume that this equality holds even at non-linear order (this assumption is numerically verified). The sector of the BdG Hamiltonian for the $g$ fermions is given by
\begin{equation}
\mathcal{H}_{mf}=\frac{1}{2}\sum_k
\begin{pmatrix}
g_\textbf{k}^\dagger & \bar{g}_\textbf{k}^\dagger & g_{-\textbf{k}} & \bar{g}_{-\textbf{k}}
\end{pmatrix}
\begin{pmatrix}
\varepsilon_g(\textbf{k}) & 0 & -\Delta_{gg}\cos\varphi_k & i\Delta_{gg}\sin\varphi_k \\
0 & \varepsilon_{\bar{g}}(\textbf{k}) & i\Delta_{gg}\sin\varphi_k & -\Delta_{gg}\cos\varphi_k \\
-\Delta_{gg}\cos\varphi_k & -i\Delta_{gg}\sin\varphi_k & -\varepsilon_g(\textbf{k}) & 0 \\
-i\Delta_{gg}\sin\varphi_k & -\Delta_{gg}\cos\varphi_k & 0 & -\varepsilon_{\bar{g}}(\textbf{k})
\end{pmatrix}
\begin{pmatrix}
g_\textbf{k} \\ \bar{g}_\textbf{k} \\ g^\dagger_{-\textbf{k}} \\ \bar{g}^\dagger_{-\textbf{k}}
\end{pmatrix}
\end{equation}
In the equation above, we have absorbed the factors $\gamma_k$ and $z_k$ into $\Delta_{gg}$ for simplicity, since those are essentially constants in the small area around the Fermi surface in which the gap is defined. The electronic excitation spectrum for this Hamiltonian is given by
\begin{equation}
E_{\pm}^2(\textbf{k})=\left(\sqrt{\xi_{+}^2(\textbf{k})+\Delta_{gg}^2\sin^2\varphi_k}\pm\Big|\xi_{-}(\textbf{k})\Big|\right)^2+\Delta_{gg}^2\cos^2\varphi_k
\end{equation}
where we have defined
\begin{equation}
\xi_{\pm}(\textbf{k})=\frac{\varepsilon_g(\textbf{k})\pm\varepsilon_{\bar{g}}(\textbf{k})}{2}
\end{equation}
One can see that nodes can develop at $E_{-}(\textbf{k})$ only along the $k_y$ axis and at momenta that satisfy the condition $\varepsilon_{{g}}(\textbf{k})\varepsilon_{\bar{g}}(\textbf{k})=-\Delta_{gg}^2$. Assuming that $\varepsilon_{g/\bar{g}}(k)=\xi_k \pm h$ for simplicity, we get the condition 
\begin{equation}
\xi_k=\pm\sqrt{h^2-\Delta_{gg}^2}
\end{equation}
for the appearance of nodes. This equation implies that two nodal points (one at $\varphi_k=\pi/2$ and a second at $\varphi_k=3\pi/2$) appear at $k_y=k_F$ at a magnetic field $h=\Delta_{gg}$, where $k_F$ is the Fermi momentum in the normal state and at zero field. Expanding the BdG lower band around the nodal points we find that the dispersion is parabolic along the $k_y$ axis. Namely, we get
\begin{equation}
E_{-}(\textbf{k})\approx\frac{v_F^2\tilde{k}_y^2}{2\Delta_{gg}}
\end{equation}
where $\tilde{k}_y$ denotes the deviation from the nodal point. For magnetic fields $h>\Delta_{gg}$, each node splits into two. Both nodal points lie between the magnetic field split normal state pockets defined by $\varepsilon_g(k)=0$ and $\varepsilon_{\bar{g}}(k)=0$. In the limit $\Delta_{gg} \to 0$, these points approach the corresponding pocket boundaries. Expanding the BdG lower band around the nodal points we now find
\begin{equation}
E_{-}(\textbf{k})\approx\frac{\sqrt{h^2-\Delta_{gg}^2}}{h}v_F\big|\tilde{k}_y\big|
\end{equation}
\end{document}